\documentclass[11pt,a4paper]{article}

\usepackage{amsmath,amsfonts,graphicx,natbib,psfrag,color}
\usepackage{algorithm, algorithmic}
\usepackage[textsize=small]{todonotes}
\usepackage{booktabs}
\usepackage{cleveref}
\usepackage[parfill]{parskip}
\usepackage{multirow}

\newcommand{\X}{\mathsf{X}}

\newcommand{\Y}{\mathsf{Y}}

\newcommand{\defeq}{\mathrel{\mathop:}=}

\title{Nested ensemble Kalman filter for static parameter inference in nonlinear state-space models}

\author{Andrew Golightly$^1$\footnote{andrew.golightly@durham.ac.uk}, Sarah E. Heaps$^1$, Chris Sherlock$^2$, Laura, E. Wadkin$^3$,\\ Darren J. Wilkinson$^1$}
\date{\small $^1$ Department of Mathematical Sciences, Durham University, UK\\ $^2$ School of Mathematical Sciences, Lancaster University, UK\\ $^3$ School of Mathematics, Statistics and Physics, Newcastle University, UK}

\begin{document}
\maketitle
\begin{abstract}
The ensemble Kalman filter (EnKF) is a popular technique for performing inference in state-space models (SSMs), particularly when the dynamic process is high-dimensional. Unlike reweighting methods such as sequential Monte Carlo (SMC, \emph{i.e.} particle filters), the EnKF leverages either the linear Gaussian structure of the SSM or an approximation thereof, to maintain diversity of the sampled latent states (the so-called ensemble members) via shifting-based updates. Joint parameter and state inference using an EnKF is typically achieved by augmenting the state vector with the static parameter. In this case, it is assumed that both parameters and states follow a linear Gaussian state-space model, which may be unreasonable in practice. In this paper, we combine the reweighting and shifting methods by replacing the particle filter used in the SMC\textsuperscript{2} algorithm of \cite{Chopin2013}, with the ensemble Kalman filter. Hence, parameter particles are weighted according to the estimated observed-data likelihood from the latest observation computed by the EnKF, and particle diversity is maintained via a resample-move step that targets the marginal parameter posterior under the EnKF. Extensions to the resulting algorithm are proposed, such as the use of a delayed acceptance kernel in the rejuvenation step and incorporation of nonlinear observation models. We illustrate the resulting methodology via several applications.        
\end{abstract}

\noindent\textbf{Keywords}: ensemble Kalman filter, particle filter, SMC$^2$, state-space models

\section{Introduction}
\label{sec:intro}

State-space models \citep[SSMs, \emph{e.g.}][]{harvey1990forecasting,shumway2006,Sarkka2013} provide a flexible framework for modelling time-series data. An SSM typically involves two components: a model describing the temporal evolution of the latent system state and a model linking the state to observations. When these components involve linear functions of the latent process, the result is a tractable state-space model, often referred to as a dynamic linear model in the discrete-time context \citep[DLM, see \emph{e.g.}][]{west2006bayesian,petris2009dynamic}. In this case, the tractable observed-data likelihood and smoothing process can be efficiently calculated (conditional on parameters) using a Kalman filter \citep{kalman1960} and joint state and parameter inference can proceed in the Bayesian context via Markov chain Monte Carlo \citep[\emph{e.g.}][]{carter1994gibbs,fruhwirth1994data,GamermanLopes06}. For nonlinear state-space models, inference may proceed via pseudo-marginal methods \citep{andrieu+r09} and, in particular, particle MCMC \citep[][]{Andrieu2010}. 

The focus of this paper is sequential Bayesian inference for the parameters and latent states in nonlinear state-space models. From a state filtering perspective, particle filters \citep[\emph{e.g.}][]{Gordon1993,Doucet2011,Jacob2015} can be used to recycle posterior samples from one time point to the next through a series of simple reweighting and resampling steps. State and parameter filtering is complicated by the exponential decrease over time in distinct parameter particles. This issue can be mitigated through the use of resample-move (also known as mutation) steps, which rejuvenate parameter particles via a Metropolis-Hastings kernel that has invariant distribution given by the parameter posterior. An example of the resulting scheme in the context of a tractable observed-data likelihood can be found in \cite{chopin2002}. A pseudo-marginal analogue of this scheme, applicable in the intractable observed-data likelihood context, weights parameter particles by a likelihood estimate obtained from a particle filter over states, and rejuvenates parameter particles via a pseudo-marginal Metropolis-Hastings kernel. The resulting algorithm is termed \emph{SMC}$^2$ \citep[][]{Chopin2013}, due to the nested use of particle filters. 

It is well known that particle filters suffer from weight degeneracy for high-dimensional targets \citep[see][for a discussion]{Agapiou2017}. Moreover, the simplest, bootstrap particle filter propagates particles myopically, irrespective of the next observation, which can lead to highly variable weights when observation noise is small relative to the noise in the state process. These problems are addressed, at the expense of some inferential accuracy, by the ensemble Kalman filter (EnKF, \citealp{evensen1994}; see also \citealp{Katzfuss2016} for an introduction), which has been ubiquitously applied in a field of research originating in the geosciences termed \emph{data assimilation} \citep[\emph{e.g.}][]{evensen2022}. As with the bootstrap particle filter, the EnKF propagates a set of simulations (known as ensemble members) via forward simulation of the state process. Unlike the approach taken by the particle filter, forward samples are shifted, rather than reweighted and subsequently resampled, by leveraging a Gaussian approximation of the state distribution at the observation time, and if need be, of the observation model. Thus, weight collapse is avoided, potentially at the expense of inferential accuracy. 

Compared to the state filtering problem, relatively few works have addressed both state and parameter filtering. Moreover, commonly used approaches are often based on strong assumptions. For example, \emph{state augmentation} \citep[\emph{e.g.}][]{anderson01,Evensen2007}, assumes a linear Gaussian relationship between parameters, observations and states, with the latter augmented to include all parameter components. Several sequential and batch inference schemes are given in \cite{Katzfuss2019} \citep[see also][]{Stroud2018} in the context of dynamic parameters. These are applicable in the static parameter context through the assumption of an artificial evolution kernel, as has also been used in the particle filter context \citep[\emph{e.g.}][]{liu2001}. However, this approach can be sensitive to the choice of tuning parameters \citep{vieira2016}.

Our contribution is a scheme for state and parameter filtering in which parameter particles are weighted by an approximate observed-data likelihood contribution, computed using the EnKF. Parameter rejuvenation mutates particles through a Metropolis-Hastings kernel that again uses the EnKF to calculate the likelihood for new parameter values. This corresponds to one iteration of the (offline) ensemble MCMC strategy of \cite{Katzfuss2019} and \cite{drovandi22}. The resulting \emph{nested ensemble Kalman filter (NEnKF)} uses the same nesting approach as SMC$^2$, albeit with the particle filter over states replaced by the EnKF. Given the robustness of the EnKF over a particle filter in high-dimensional settings, we anticipate that the NEnKF will be of most benefit to practitioners working with models in which the number of static parameter components is small relative to the dimension of the hidden state process. Several examples of such a scenario are given in \cite{Katzfuss2019}. 

As with SMC$^2$, the resample-move step in the NEnKF is the main computational bottleneck. This can be alleviated at the expensive of inferential accuracy, for example, by calculating the likelihood over a moving window rather than the full time horizon \cite[\emph{e.g.}][]{delmoral2017biased} or by replacing the resample-move step with parameter particles sampled from a jittering kernel \cite[\emph{e.g.}][]{crisan18,fang2024}. In this work, we choose to improve the efficiency of the resample-move through the use of a delayed-acceptance kernel \cite[\emph{e.g.}][]{christen2005,Goli15,Banterle19} that first tries proposals using a simple $k$-nearest neighbour approximation to the observed data likelihood, that can be constructed dynamically using the current particle set. 

We also consider the case of nonlinear and non-Gaussian observation models, for which using the EnKF directly can lead to biased or inaccurate estimates of the filtering distribution \citep[\emph{e.g.}][]{zhang2010,saetrom2011,grooms2022}. In this case we propose to use the EnKF to propagate the state process inside the inner particle filter of an SMC$^2$ scheme, so that the bias in the enKF is corrected for via an appropriate weight function. Unlike related approaches which use the EnKF as a proposal mechanism inside a particle filter \citep[\emph{e.g.}][]{bi15,tamboli2021}, our novel method allows for marginalising the latent state process between observation times.

Finally, we note that replacing the inner particle filter of SMC$^2$ with the ensemble Kalman filter has been proposed independently of our work by \cite{temfack25}. Unlike our approach, the main focus of their work is sequential inference for stochastic epidemic models, with the EnKF adapted to epidemic data by incorporating state-dependent observation variance, as is necessary for over-dispersed incidence counts. Additionally, and as in \cite{drovandi22}, an unbiased Gaussian density estimator is used to mitigate finite-sample effects. The nested use of the EnKF in \cite{temfack25} further underlines the possible computational gains of using the EnKF versus a vanilla SMC$^2$ implementation.

The remainder of this paper is organised as follows. Section~\ref{sec:bg} introduces necessary background material on state-space models, the particle filter and SMC$^2$. Section~\ref{sec:KFinf} considers existing approaches to inference based on the EnKF. Section~\ref{sec:nest} outlines our proposed approach to the filtering problem and several extensions thereof. Applications of the proposed methodology are considered in Section~\ref{sec:app} and conclusions drawn in Section~\ref{sec:disc}.

\section{Background}
\label{sec:bg}

This section describes the relevant preliminary material. In particular, we consider state-space models and the inference objective in Section~\ref{subsec:ssm}, the particle filter is detailed in Section~\ref{subsec:pf} and its use inside SMC$^2$ is described in Section~\ref{subsec:smc2}. 

\subsection{State-space models and inference task} \label{subsec:ssm}

A state-space model is defined for a fixed parameter vector $\theta=(\theta_1,\ldots,\theta_{d_{\theta}})\in\Theta$ by the hidden / latent state dynamics $X_{0:t}\defeq (X_0,\ldots, X_t)$, $X_t\in \X$, given by initial distribution
$
X_0 \sim p_0^{(\theta)}(\cdot),
$
and 
\begin{align*}
  X_t &\sim p_t^{(\theta)}(\cdot|x_{t-1}),
  \qquad t=1,\ldots, T,
  \\
  Y_t &\sim f_t^{(\theta)}(\cdot|x_{t}),\hspace{0.3cm}
  \qquad t=0,\ldots, T.
\end{align*}
Here, $p^{(\theta)}_t(\cdot|x)$ and $f^{(\theta)}(\cdot|x)$ are conditional probability densities given $x$, and $Y_{0:T}$, $Y_t\in \Y$,  is the resulting sequence of observations. Unless stated otherwise, we will assume that each $X_t$ and $Y_t$ are random vectors with support $\X=\mathbb{R}^{d_x}$ and $\Y=\mathbb{R}^{d_y}$ respectively.  

The joint density of the hidden states and observations is given by
 \[
 p_T^{(\theta)}(x_{0:T},y_{0:T})
 = \left\{ \prod_{t=0}^T  f_t^{(\theta)}(y_t|x_t)\right\} p_0^{(\theta)}(x_0) \prod_{t=1}^T p_t^{(\theta)}(x_t|x_{t-1}). 
 \]
Marginalising out the hidden states gives the \emph{observed-data likelihood}
\[
p_T^{(\theta)}(y_{0:T})=p_0^{(\theta)}(y_0)\prod_{t=1}^T p_t^{(\theta)}(y_t|y_{0:t-1}).
\]
From a Bayesian perspective, inference for the static parameters may proceed via the marginal parameter posterior
\begin{equation}\label{ppost}
\pi_T(\theta|y_{0:T})\propto \pi(\theta)p_T^{(\theta)}(y_{0:T})
\end{equation}
where $\pi(\theta)$ is the prior density ascribed to $\theta$. If desired, samples of the hidden states can be generated from the smoothing density $p_T^{(\theta)}(x_{0:T}|y_{0:T})$, which is proportional to $p_{T}^{(\theta)}(x_{0:T},y_{0:T})$.  


In this article, the goal is sequential inference via recursive use of  filtering densities of the form
\begin{align}\label{jointfilt}
\pi_t(\theta,x_{t}|y_{0:t}) &= 
\pi_{t}(\theta|y_{0:t})p_t^{(\theta)}(x_{t}|y_{0:t}),\quad t=0,1,\ldots,T.
\end{align}
For a fixed and known parameter $\theta$, recursive exploration of the filtering densities over hidden states, $p_t^{(\theta)}(x_{t}|y_{0:t})$, $t=0,1,\ldots,T$, is typically achieved via a particle filter, which we consider in the next section. 


\subsection{Particle filter}\label{subsec:pf}
The bootstrap particle filter \citep[see \emph{e.g.}][]{Gordon1993} consists of a sequence of weighted resampling steps, whereby upon receipt of $y_t$, $N$ state \emph{particles} $x_{t-1}^{1:N}=\{x_{t-1}^{1},\ldots,x_{t-1}^{N}\}$ are propagated forward via some proposal density, appropriately weighted and resampled with replacement. 

The particle filter operates recursively as follows. The filtering density at time $t$ can be written as
\begin{equation}\label{filt}
p_t^{(\theta)}(x_{t}|y_{0:t})\propto f_t^{(\theta)}(y_t|x_t)\int p_t^{(\theta)}(x_t|x_{t-1}) p_{t-1}^{(\theta)}(x_{t-1}|y_{0:t-1})dx_{t-1}
\end{equation}
which is typically intractable. A weighted sample $\{x_{t-1}^{1:N},w_{t-1}^{1:N}\}$ from $p_{t-1}^{(\theta)}(x_{t-1}|y_{0:t-1})$ admits the approximation 
\[
\widehat{p}_t^{(\theta)}(x_{t}|y_{0:t})\propto f_t^{(\theta)}(y_t|x_t)\sum_{j=1}^N  p_t^{(\theta)}(x_t|x_{t-1}^{j})w_{t-1}^{j},
\]
which is sampled by first drawing an $x_{t-1}^j$ with probability $w_{t-1}^j$ then propagating according to a proposal density $q_t^{(\theta)}(x_t|x_{t-1},y_t)$ and finally, reweighting accordingly. This weighted resampling scheme is presented in Algorithm~\ref{alg:PartF}, recursive application of which corresponds to the particle filter. Note that at time $t=0$, steps 2--3 are implemented such that the propagate step draws $x_0^{j}\sim p_0^{(\theta)}(\cdot)$ and then the (unnormalised) weight is computed as $f_0^{(\theta)}(y_0|x_0^{j})$.    

\begin{algorithm}[htb] \caption{Particle filter (step $t$)} \label{alg:PartF}
\begin{algorithmic}
\STATE Input: parameter $\theta$ and weighted particles $\{x_{t-1}^{1:N},w_{t-1}^{1:N}\}$.
\FOR{$j=1,2,\ldots,N$}
\STATE \textbf{1. Resample}. Sample an index $a_{t-1}^j$ from $\{1,\ldots,N\}$ with probabilities $w_{t-1}^{1:N}$. 
\STATE \textbf{2. Propagate}. Sample ${x}_t^{j}\sim q_t^{(\theta)}(\cdot|x_{t-1}^{a_{t-1}^j},y_t)$.
\STATE \textbf{3. Weight}. Construct and normalise the weight 
\[
\tilde{w}_{t}(x_{t-1}^{a_{t-1}^j},x_t^{j})=\frac{p_t^{(\theta)}(x_t^{j}|x_{t-1}^{a_{t-1}^j})f_t^{(\theta)}(y_t|x_t^{j})}{q_t^{(\theta)}(x_{t}^{j}|x_{t-1}^{a_{t-1}^j},y_t)}, \quad w_t^{j}=\tilde{w}_{t}(x_{t-1}^{a_{t-1}^j},x_t^{j})/\sum_{k=1}^{N}\tilde{w}_{t}(x_{t-1}^{a_{t-1}^k},x_t^{k})
.\] 
\ENDFOR
\STATE Output: likelihood estimate $\widehat{p}_{t}^{(\theta)}(y_t|y_{0:t-1},x_t^{1:N},a_{t-1}^{1:N}) = \frac{1}{N}\sum_{j=1}^N \tilde{w}_{t}(x_{t-1}^{a_{t-1}^j},x_t^{j})$ and weighted particles $\{x_{t}^{1:N},w_t^{1:N}\}$.
\end{algorithmic}
\end{algorithm}

The particle filter can be used to unbiasedly estimate the observed-data likelihood via the product (over time) of the average unnormalised weights  \citep[see \emph{e.g.}][]{DelMoral2004,pitt12}. The estimator is
\begin{align}\label{llikeEst}
\widehat{p}_T^{(\theta)}(y_{0:T}|x_{0:T}^{1:N},a_{0:T-1}^{1:N})&=
\widehat{p}_0^{(\theta)}(y_0|x_0^{1:N})\times \prod_{t=1}^T \widehat{p}_t^{(\theta)}(y_t|y_{0:t-1},x_{t}^{1:N},a_{t-1}^{1:N})\nonumber\\
&=N^{-T-1}\sum_{j=1}^N \tilde{w}_0(x_0^{j})\times \prod_{t=1}^T \left\{\sum_{j=1}^N \tilde{w}_t(x_{t-1}^{a_{t-1}^j},x_t^j) \right\}
\end{align}
which we let depend explicitly on the particles $x_{0:T}^{1:N}$ and their genealogy $a_{0:T-1}^{1:N}$. In the context of batch inference for the static parameters $\theta$, (\ref{llikeEst}) can be used inside a pseudo-marginal Metropolis-Hastings (PMMH) scheme which exactly targets $\pi_T(\theta|y_{0:T})$ \citep[see \emph{e.g.}][]{Andrieu2010}. The contributions in (\ref{llikeEst}) can also be used as incremental weights in a particle filter over static parameters, which we briefly review in the next section.   

\subsection{SMC$^2$}\label{subsec:smc2}
Suppose now that $\theta$ is unknown and interest lies in recursive learning of $\pi_t(\theta|y_{0:t})$, $t=0,\ldots,T$. Sequential use of Bayes Theorem gives
\begin{equation}\label{postSMC2}
\pi_{t}(\theta|y_{0:t})\propto \pi_{t-1}(\theta|y_{0:t-1})p_{t}^{(\theta)}(y_t|y_{0:t-1})
\end{equation}
which is complicated by the observed-data likelihood $p_{t}^{(\theta)}(y_t|y_{0:t-1})$, which is typically unavailable in closed form. Given a weighted sample of size $M$, $\{\theta^{1:M},u_{t-1}^{1:M}\}$, 
from $\pi_{t-1}(\theta|y_{0:t-1})$, an idealised SMC scheme generates a weighted sample distributed according to (\ref{postSMC2}) by reweighting each $\theta^i$ according to  $u_t^i\propto u_{t-1}^i p_{t}^{(\theta^i)}(y_t|y_{0:t-1})$. Triggered by the fulfilment of some degeneracy criterion, such as that related to (\ref{ess}), below, a resample-move step \citep[see \emph{e.g.}][]{gilks2001} generates new parameter particles via a Metropolis-Hastings kernel that has (\ref{postSMC2}) as its invariant distribution. Full details of the resulting iterated batch importance sampling (IBIS) scheme are given in \cite{chopin2002}. 

We consider here the pseudo-marginal analogue of IBIS \citep[termed \emph{SMC}$^2$,][]{Chopin2013}, wherein each $\theta^i$ particle is reweighted according to $u_t^i\propto u_{t-1}^i \widehat{p}_{t}^{(\theta^i)}(y_t|y_{0:t-1},x_{t}^{i,1:N},a_{t-1}^{i,1:N})$, and $\widehat{p}_{t}^{(\theta^i)}(y_t|y_{0:t-1},x_{t}^{i,1:N},a_{t-1}^{i,1:N})$ is the estimate of observed-data likelihood obtained by running Algorithm~\ref{alg:PartF} with parameter $\theta^i$ and the associated state particle system $\{x_{t-1}^{i,1:N},w_{t-1}^{i,1:N}\}$. Hence, over the entire observation period, each particle $\theta^i$ has associated state particles and genealogies $\{x_{0:T}^{i,1:N},a_{0:T-1}^{i,1:N}\}$. 

Parameter weight degeneracy is monitored via effective sample size (ESS), which is calculated at time $t$ using the normalised parameter particle weights as
\begin{equation}\label{ess}
\textrm{ESS}=1/\sum_{i=1}^M (u_t^{i})^2.
\end{equation}
If ESS falls below some user specified threshold, $\gamma M$, a resample-move step is triggered by drawing $M$ times from the mixture $\sum_{i=1}^M u_t^{i} K_t\left(\theta^{i}\,,\, \cdot \right)$, 
where $K_t$ is a pseudo-marginal Metropolis-Hastings kernel with target $\pi_{t}(\theta|y_{0:t})$. Hence, after resampling, a new particle $\theta^*$ is proposed from a kernel $q(\theta^{i},\cdot)$ and accepted with probability 
\begin{equation}\label{aprobSMC2}
\alpha(\theta^*|\theta^{i}) = 1 \wedge \frac{\pi(\theta^*)}{\pi(\theta^{i})}\times \frac{\widehat{p}_{t}^{(\theta^*)}(y_{0:t}|x_{0:t}^{*,1:N},a_{0:t-1}^{*,1:N})}{\widehat{p}_{t}^{(\theta^{i})}(y_{0:t}|x_{0:t}^{i,1:N},a_{0:t-1}^{i,1:N})}\times    
\frac{q(\theta^*,\theta^{i})}{q(\theta^{i},\theta^*)}.
\end{equation}
A formal justification of SMC$^2$ can be found in \cite{Chopin2013}; briefly, if $\psi_t^{(\theta)}(x_{0:t}^{1:N},a_{0:t-1}^{1:N})$ denotes the joint probability density of all random variables generated by the state particle filter up to time $t$, then SMC$^2$ recursively targets
\[
\widehat{\pi}_{t}(\theta,x_{0:t}^{1:N},a_{0:t-1}^{1:N}|y_{0:t})\propto \pi(\theta) \widehat{p}_{t}^{(\theta)}(y_{0:t}|x_{0:t}^{1:N},a_{0:t-1}^{1:N}) \psi_t^{(\theta)}(x_{0:t}^{1:N},a_{0:t-1}^{1:N})
\]
which admits $\pi_t(\theta|y_{0:t})$ as a marginal. 

\begin{algorithm}[htb] \caption{SMC$^2$ (step $t$)} \label{alg:smc2}
\begin{algorithmic}
\STATE Input: weighted parameter particles $\{\theta^{i},u_{t-1}^{i}\}$, associated weighted particles $\{x_{t-1}^{i,1:N},w_{t-1}^{i,1:N}\}$ and likelihood evaluations $\widehat{p}_{t-1}^{(\theta^{i})}(y_{0:t-1}|x_{0:t-1}^{i,1:N},a_{0:t-2}^{i,1:N})$, $i=1,\ldots,M$.
\FOR{$i=1,2,\ldots,M$}
\STATE \textbf{1. Update weights and likelihood}. Run Algorithm~\ref{alg:PartF} with inputs $\theta^{i}$ and $\{x_{t-1}^{i,1:N},w_{t-1}^{i,1:N}\}$ to obtain $\widehat{p}_{t}^{(\theta^{i})}(y_t|y_{0:t-1},x_t^{i,1:N},a_{t-1}^{i,1:N})$ and $\{x_{t}^{i,1:N},w_t^{i,1:N}\}$. Construct the weight 
\[
\tilde{u}_{t}^{i}=u_{t-1}^{i}\widehat{p}_{t}^{(\theta^{i})}(y_t|y_{0:t-1},x_{t}^{i,1:N},a_{t-1}^{i,1:N})
\]
and update the likelihood 
\[
\widehat{p}_{t}^{(\theta^{i})}(y_{0:t}|x_{0:t}^{i,1:N},a_{0:t-1}^{i,1:N})=\widehat{p}_{t-1}^{(\theta^{i})}(y_{0:t-1}|x_{0:t-1}^{i,1:N},a_{0:t-2}^{i,1:N})\widehat{p}_{t}^{(\theta^{i})}(y_t|y_{0:t-1},x_t^{i,1:N},a_{t-1}^{i,1:N}).
\]
\IF{$\textrm{ESS} < \gamma M$}
\STATE \textbf{2. Resample}. 
Normalise the weights to give $u_t^{i}=\tilde{u}_t^{i}/\sum_{k=1}^M \tilde{u}_t^{k}$. 
 Sample an index $b_i$ from $\{1,\ldots,M\}$ with probabilities $u_t^{1:M}$. 
 Put $\{\theta^{i},u_t^{i}\}:=\{\theta^{b_i},1/M\}$ and $\{x_{t}^{i,1:N},w_{t}^{i,1:N}\}:=\{x_{t}^{b_i,1:N},w_{t}^{b_i,1:N}\}$.
 \STATE \textbf{3. Move / mutate}. Propose $\theta^*\sim q(\theta^{i},\cdot)$. Perform iterations $1,\ldots,t$ of Algorithm~\ref{alg:PartF} to obtain $\widehat{p}_{t}^{(\theta^{*})}(y_{0:t}|x_{0:t}^{*,1:N},a_{0:t-1}^{*,1:N})$ and $\{x_{t}^{*,1:N},w_{t}^{*,1:N}\}$. With probability~(\ref{aprobSMC2}), put $\theta^{i}=\theta^*$, $\widehat{p}_{t}^{(\theta^{i})}(y_{0:t}|x_{0:t}^{i,1:N},a_{0:t-1}^{i,1:N})=\widehat{p}_{t}^{(\theta^{*})}(y_{0:t}|x_{0:t}^{*,1:N},a_{0:t-1}^{*,1:N})$ and $\{x_{t}^{i,1:N},w_{t}^{i,1:N}\}=\{x_{t}^{*,1:N},w_{t}^{*,1:N}\}$.
\ENDIF
\ENDFOR
\STATE Output: weighted parameter particles 
$\{\theta^{i},u_{t}^{i}\}$, associated weighted particles $\{x_{t}^{i,1:N},w_{t}^{i,1:N}\}$ and likelihood evaluations $\widehat{p}_{t}^{(\theta^{i})}(y_{0:t}|x_{0:t}^{i,1:N},a_{0:t-1}^{i,1:N})$, $i=1,\ldots,M$.
\end{algorithmic}
\end{algorithm}

Step $t$ of SMC$^2$ is given in Algorithm~\ref{alg:smc2} and assumes a fixed number of state particles, $N$. Since our goal is filtering via (\ref{jointfilt}), we only store $\{\theta^i,u_t^i\}$ and the associated weighted particles $\{x_{t}^{i,1:N},w_{t}^{i,1:N}\}$ at the most recent time, giving a storage cost of $\mathcal{O}(MN)$. Conditional on $\theta^i$, a draw from $p^{(\theta^i)}(x_t|y_{0:t})$ can be obtained by generating an index $n(i)$ from $\{1,\ldots,N\}$ with probabilities $w_t^{i,1:N}$ and returning $x_t^{i,n(i)}$. 

Note that the overall acceptance rate of the mutation step will depend, \emph{inter alia}, on the variance of the likelihood approximation, which can be controlled by dynamically choosing the number of state particles $N$, for example, by doubling $N$ if the empirical acceptance rate falls below some threshold chosen by the practitioner. For a given $\theta$ particle, state particles and their genealogies $\{x_{0:t}^{1:N},a_{0:t-1}^{1:N}\}$ are replaced by $\{\check{x}_{0:t}^{1:\check{N}},\check{a}_{0:t-1}^{1:\check{N}}\}$ (generated through recursive application of Algorithm~\ref{alg:PartF}) using the generalised importance sampling strategy of \cite{delMoral06}. Implementation of this step implicitly weights each $\theta^i$ particle by
\[
\frac{\widehat{\pi}_{t}(\theta^i,\check{x}_{0:t}^{i,1:\check{N}},\check{a}_{0:t-1}^{i,1:\check{N}}|y_{0:t}) L_t^{(\theta^i)}\big(\{\check{x}_{0:t}^{i,1:\check{N}},\check{a}_{0:t-1}^{i,1:\check{N}}\},\{x_{0:t}^{1:N},a_{0:t-1}^{i,1:N}\}\big)}{\widehat{\pi}_{t}(\theta^i,x_{0:t}^{i,1:N},a_{0:t-1}^{i,1:N}|y_{0:t})\psi_t^{(\theta)}(\check{x}_{0:t}^{i,1:\check{N}},\check{a}_{0:t-1}^{i,1:\check{N}}) }
\]
where $L_t^{(\theta)}$ is an artificial backward kernel. As discussed in \cite{Chopin2013}, although the optimal backward kernel is typically intractable, $L_t^{(\theta)}$ can be chosen to give a simple incremental weight of the form $\widehat{p}_{t}^{(\theta^i)}(y_{0:t}|\check{x}_{0:t}^{i,1:\check{N}},\check{a}_{0:t-1}^{i,1:\check{N}})/\widehat{p}_{t}^{(\theta^i)}(y_{0:t}|x_{0:t}^{i,1:N},a_{0:t-1}^{i,1:N})$. On the other hand, \cite{botha23} suggest a choice of backward kernel that introduces no additional variability in the particle weights; the incremental weight is 1 in this case. 

\section{Inference using the ensemble Kalman filter}\label{sec:KFinf}
In this section, we briefly review several existing methods for sequential state and parameter inference based on the ensemble Kalman filter (EnKF). For a detailed discussion, we refer the reader to \cite{Katzfuss2019}.

\subsection{Ensemble Kalman Filter} \label{subsec:enkf}

The ensemble Kalman filter \citep[EnKF, see \emph{e.g.}][]{evensen1994,Katzfuss2016} can be used to give approximate draws from the filtering densities $p_t^{(\theta)}(x_{t}|y_{0:t})$, $t=0,1,\ldots,T$.  We assume here a linear and Gaussian observation model; extensions of the EnKF to other observation models are considered in Section~\ref{NGobs}. 

Hence, consider
\begin{equation} \label{eq:gauss}
f_t^{(\theta)}(y_t|x_t)= \mathcal{N}(y_t; H x_t, R)
\end{equation}
where $\mathcal{N}(\cdot;\mu,\Sigma)$ denotes the density of a Gaussian random vector with mean $\mu$ and variance $\Sigma$. Additionally, $H$ is a $d_y \times d_x$ matrix and $R$ is a $d_y\times d_y$ variance matrix, and both of these may be functions of $\theta$. The matrix $H$ determines the observation regime; for example, taking $H = I$, the $d_x \times d_x$ identity matrix, gives a complete observation regime in which all elements of $X_t$ are observed subject to Gaussian error. We assume throughout that $H$ and $R$ are constant, although the EnKF can accommodate time dependence of these quantities.

The EnKF operates recursively as follows. Suppose that a sample of size $N$ \emph{ensemble members} $x_{t-1}^{1:N}=\{x_{t-1}^{1},\ldots,x_{t-1}^{N}\}$ is available at time $t-1$ from $p_{t-1}^{(\theta)}(x_{t-1}|y_{0:t-1})$. A forecast ensemble of size $N$ is generated from $p_{t}^{(\theta)}(x_t|y_{0:t-1})$ by sampling $\tilde{x}_t^{j}\sim p_t^{(\theta)}(\cdot|x_{t-1}^{j})$, $j=1,\ldots,N$, to give $\tilde{x}_{t}^{1:N}$. Key to the operation of the EnKF is to approximate the forecast density as
\begin{equation}\label{eq:fore}
\widehat{p}_{\textrm{enkf},t}^{(\theta)}(x_t|y_{0:t-1},\tilde{x}_t^{1:N})=\mathcal{N}(x_t\,;\,\hat{\mu}_{t|t-1}\,,\,\hat{\Sigma}_{t|t-1})
\end{equation}
where $\hat{\mu}_{t|t-1}$ and $\hat{\Sigma}_{t|t-1}$ are typically taken to be the sample mean and variance of the forecast ensemble $\tilde{x}_t^{1:N}$. Then, Bayes Theorem gives the approximation to the filtering density based on the EnKF as
\begin{equation}\label{eq:filt}
\widehat{p}_{\textrm{enkf},t}^{(\theta)}(x_t|y_{0:t},\tilde{x}_t^{1:N})= \frac{\widehat{p}_{\textrm{enkf},t}^{(\theta)}(x_t|y_{0:t-1},\tilde{x}_t^{1:N})f_t^{(\theta)}(y_t|x_t)}{\widehat{p}_{\textrm{enkf},t}^{(\theta)}(y_t|y_{0:t-1},\tilde{x}_t^{1:N})}
\end{equation}
for which, under the linear Gaussian structure of (\ref{eq:gauss}) and (\ref{eq:fore}), we obtain
\begin{equation}\label{enkfPost}
\widehat{p}_{\textrm{enkf},t}^{(\theta)}(x_t|y_{0:t},\tilde{x}_t^{1:N})=\mathcal{N}(x_t\,;\,\hat{\mu}_{t|t}\,,\,\hat{\Sigma}_{t|t})    
\end{equation}
where $\hat{\mu}_{t|t}=\hat{\mu}_{t|t-1}+\hat{K}_t(y_t-H\hat{\mu}_{t|t-1})$, $\hat{\Sigma}_{t|t}=(I-\hat{K}_t H)\hat{\Sigma}_{t|t-1}$ 
and $\hat{K}_t$ is an estimate of the Kalman gain, that is
\begin{equation}\label{Kgain}
\hat{K}_t = \hat{\Sigma}_{t|t-1}H'(H\hat{\Sigma}_{t|t-1} H'+R)^{-1}.
\end{equation}
Although (\ref{enkfPost}) can be sampled from trivially, it is common to perform a stochastic update step \cite[see e.g.][]{Katzfuss2016}, which moves ensemble members according to a Gaussian pseudo-observation $\tilde{y}_t^{j}\sim \mathcal{N}(H\tilde{x}_t^{j},R)$. New ensemble members are obtained as $x_{t}^{j}=\tilde{x}_t^{j}+\hat{K}_t(y_t-\tilde{y}_t^{j})$ which avoids draws of dimension $d_x$ (with $d_x >> d_y$ in common applications). Note that marginalising over $x_t$ in (\ref{eq:gauss}) with respect to (\ref{eq:fore}) gives the denominator in (\ref{eq:filt}) (that is, observed data likelihood contribution) as
\begin{equation}\label{likeInc}
\widehat{p}_{\textrm{enkf},t}^{(\theta)}(y_t|y_{0:t-1},\tilde{x}_t^{1:N})=\mathcal{N}(y_t\,;\,H\hat{\mu}_{t|t-1}\,,\, H\hat{\Sigma}_{t|t-1} H'+R).    
\end{equation}

Algorithm~\ref{alg:EnKF} gives a summary of the above steps upon receipt of an observation $y_t$. The EnKF can be initialised at time $t=0$ by sampling $N$ ensemble members $\tilde{x}_0^{1:N}$ from $p_0^{(\theta)}(\cdot)$ and setting $\hat{\mu}_0$ and $\hat{\Sigma}_0$ to be the sample mean and variance of $\tilde{x}_0^{1:N}$. Equations (\ref{enkfPost})--(\ref{likeInc}) are applied with $\hat{\mu}_0$ and $\hat{\Sigma}_0$ replacing $\hat{\mu}_{t|t-1}$ and $\hat{\Sigma}_{t|t-1}$  

Up until now, our discussion of the EnKF has assumed that the parameter $\theta$ is fixed and known. In the following three sections, we briefly review some existing methods for dealing with an unknown $\theta$. 

\begin{algorithm}[htb] \caption{Ensemble Kalman filter (step $t$)} \label{alg:EnKF}
\begin{algorithmic}
\STATE Input: parameter $\theta$, ensemble members $x_{t-1}^{1:N}$.
\FOR{$j=1,2,\ldots,N$}
\STATE \textbf{1. Forecast ensemble}. Sample $\tilde{x}_t^{j}\sim p_t^{(\theta)}(\cdot|x_{t-1}^{j})$.
\STATE \textbf{2. Update step}. Set $x_{t}^{j}=\tilde{x}_t^{j}+\hat{K}_t(y_t-\tilde{y}_t^{j})$, where $\tilde{y}_t^{j}\sim \mathcal{N}(H\tilde{x}_t^{j},R)$ is a pseudo-observation.
\ENDFOR
\STATE Output: likelihood increment $\widehat{p}_{\textrm{enkf},t}^{(\theta)}(y_t|y_{0:t-1},\tilde{x}_t^{1:N}) = \mathcal{N}(y_t\,;\,H\hat{\mu}_{t|t-1}\,,\, H\hat{\Sigma}_{t|t-1} H'+R)$ and ensemble members $x_{t}^{1:N}$.
\end{algorithmic}
\end{algorithm}


\subsection{Ensemble MCMC}\label{subsec:eMCMC}
Recall the marginal parameter posterior $\pi_T(\theta|y_{0:T})$ given by (\ref{ppost}). This can be approximated using the EnKF by considering the joint density
\begin{align}
\widehat{\pi}_{\textrm{enkf},T}(\theta,\tilde{x}_{0:T}^{1:N},x_{0:T}^{1:N}|y_{0:T})&\propto \pi(\theta)\widehat{p}_{\textrm{enkf},T}^{(\theta)}(y_{0:T}|\tilde{x}_{0:T}^{1:N})\psi_T^{(\theta)}(\tilde{x}_{0:T}^{1:N},x_{0:T}^{1:N})\nonumber\\
&\propto \pi(\theta)\left\{\widehat{p}_{\textrm{enkf},0}^{(\theta)}(y_0|\tilde{x}_0^{1:N})\prod_{t=1}^T \widehat{p}_{\textrm{enkf},t}^{(\theta)}(y_t|y_{0:t-1},\tilde{x}_t^{1:N})\right\} \psi_T^{(\theta)}(\tilde{x}_{0:T}^{1:N},x_{0:T}^{1:N})\label{epost}
\end{align}
where $\psi_T^{(\theta)}(\tilde{x}_{0:T}^{1:N},x_{0:T}^{1:N})$ denotes the joint density of the entire ensemble $\{\tilde{x}_{0:T}^{1:N},x_{0:T}^{1:N}\}$. Although (\ref{epost}) is typically intractable, it can be sampled using MCMC, and the parameter samples retained to give (dependent) draws from $\widehat{\pi}_{\textrm{enkf},T}(\theta|y_{0:T})$, the $\theta$-marginal of (\ref{epost}). The resulting approach, termed \emph{ensemble MCMC} (eMCMC) is detailed in \cite{drovandi22} \citep[see also][]{Stroud2018,Katzfuss2019}. In brief, at iteration $k$, a new $\theta^*$ is proposed from a kernel $q(\theta^{k-1},\cdot)$. Then, sampling of the corresponding ensemble $\{\tilde{x}_{0:T}^{*,1:N},x_{0:T}^{*,1:N}\}$ from $\psi_T^{(\theta^*)}(\cdot,\cdot)$ and evaluation of the observed data likelihood $\widehat{p}_{\textrm{enkf},T}^{(\theta^*)}(y_{0:T}|\tilde{x}_{0:T}^{*,1:N})$, is achieved through recursive application of Algorithm~\ref{alg:EnKF}. The parameter $\theta^*$ is retained with probability
\begin{equation}\label{aprobEmcmc}
\alpha_{\textrm{enkf}}(\theta^*|\theta^{k-1}) = 1 \wedge \frac{\pi(\theta^*)}{\pi(\theta^{k-1})}\times \frac{\widehat{p}_{\textrm{enkf},T}^{(\theta^*)}(y_{0:T}|\tilde{x}_{0:T}^{*,1:N})}{\widehat{p}_{\textrm{enkf},T}^{(\theta^{k-1})}(y_{0:T}|\tilde{x}_{0:T}^{k-1,1:N})}\times    
\frac{q(\theta^*,\theta^{k-1})}{q(\theta^{k-1},\theta^*)}.
\end{equation}

When applied sequentially, eMCMC is wasteful in terms of both storage and computational costs; upon receipt of a new observation (say $y_{T+1}$), previous parameter samples are discarded and eMCMC is run from scratch. Nevertheless, eMCMC is a key component of the resample-move step described in Section~\ref{sec:nest}. 


\subsection{State augmentation}\label{subsec:aug}
The EnKF can be directly modified to simultaneously estimate both hidden states and static parameters by augmenting the state vector $x_t$ to incorporate $\theta$ \citep[see e.g.][]{Evensen2007,mitchell21}. We write $z_t=(x_t',\theta_t')'$ with $x_t$ evolving according to $p_t^{(\theta)}(\cdot|x_{t-1})$ and $\theta_t=\theta_{t-1}=\theta$. The observation model becomes $f_t^{(\theta)}(y_t|z_t)= \mathcal{N}(y_t; H_z z_t, R)$ with $H_z=(H,0_{d_y \times d_{\theta}})$, where $0_{d_y \times d_{\theta}}$ is a $d_y \times d_{\theta}$ matrix of zeros and $H$ is not a function of $\theta$. 

The resulting augmented EnKF (AEnKF) operates recursively as follows. Suppose that a sample of size $N$ ensemble members 
$\{x_{t-1}^{1:N},\theta^{1:N}\}$ is 
available at time $t-1$ from the filtering distribution at time $t-1$. A forecast ensemble of size $N$ is generated via $\tilde{x}_t^{j}\sim p_t^{(\theta)}(\cdot|x_{t-1}^{j})$, $\tilde{\theta}_t=\theta_{t-1}$ and setting $\tilde{z}_t^j=(\tilde{x}_{t}^{j}{}',\tilde{\theta}_{t}^{j}{}')'$, $j=1,\ldots,N$. The update step is given by $z_{t}^{j}=\tilde{z}_t^{j}+\hat{K}_{z,t}(y_t-\tilde{y}_t^{j})$ where $\tilde{y}_t^{j}\sim \mathcal{N}(H_z\tilde{z}_t^{j},R)$ is a pseudo-observation and the Kalman gain becomes
\begin{equation}\label{Kgainz}
\hat{K}_{z,t} = \hat{\Sigma}_{t|t-1}H_z'(H_z\hat{\Sigma}_{t|t-1} H_z'+R)^{-1}.
\end{equation}
where $\hat{\Sigma}_{t|t-1}$ is the sample covariance of the forecast ensemble $\tilde{z}_t^1,\ldots,\tilde{z}_t^N$.

To alleviate under-dispersion in the parameter ensemble, for example when small ensemble sizes are considered \citep[\emph{e.g.}][]{Ruckstuhl18}, the parameter vector can be allowed to dynamically evolve, for example using the artificial evolution kernel of \cite{liu2001}, which gives
\begin{equation}\label{liu}
p_t(\theta_t|\theta_{t-1})= \mathcal{N}\left(\theta_t; a\theta_{t-1}+(1-a)\bar{\theta}_{t-1}, h^2 V_{t-1}\right)
\end{equation}
where $\bar{\theta}_{t-1}$ and $V_{t-1}$ are the sample mean and variance of $\theta_{t-1}^{1},\ldots,\theta_{t-1}^{N}$. Setting $a=\sqrt{1-h^2}$ corrects for over-dispersion; the practitioner can then choose the smoothing parameter $h$, for example using the discounting method of \cite{liu2001}. Step $t$ of the augmented EnKF is summarised by Algorithm~\ref{alg:AEnKF}. 

\begin{algorithm}[htb] \caption{Augmented ensemble Kalman filter (step $t$)} \label{alg:AEnKF}
\begin{algorithmic}
\STATE Input: ensemble members $\{x_{t-1}^{1:N},\theta_{t-1}^{1:N}\}$.
\FOR{$j=1,2,\ldots,N$}
\STATE \textbf{1. Forecast ensemble}. Sample $\tilde{\theta}_t^{j}\sim \mathcal{N}\left(a\theta_{t-1}^{j}+(1-a)\bar{\theta}_{t-1}, h^2 V_{t-1}\right)$ and 
$\tilde{x}_t^{j}\sim p_t^{(\theta_{t}^{j})}(\cdot|x_{t-1}^{j})$. Set $\tilde{z}_t^{j}=(\tilde{x}_{t}^{j}{}',\tilde{\theta}_{t}^{j}{}')'$.
\STATE \textbf{2. Update step}. Set $z_{t}^{j}=\tilde{z}_t^{j}+\hat{K}_{z,t}(y_t-\tilde{y}_t^{j})$, where $\tilde{y}_t^{j}\sim \mathcal{N}(H_z\tilde{z}_t^{j},R)$ is a pseudo-observation.
\ENDFOR
\STATE Output: ensemble members $\{x_{t}^{1:N},\theta_t^{1:N}\}$.
\end{algorithmic}
\end{algorithm}

\subsection{Particle ensemble Kalman filter (PEnKF)}\label{subsec:part}

The augmented EnKF makes the strong assumption that the parameters and observations follow a linear Gaussian structure. An alternative approach for dealing with the static parameters is to marginalise out the hidden states in a sequential fashion. Recall the factorisation of  $\pi_{t}(\theta|y_{0:t})$ in (\ref{postSMC2}) and the assumption of 
a weighted sample of size $M$, $\{\theta^{1:M},u_{t-1}^{1:M}\}$, 
from $\pi_{t-1}(\theta|y_{0:t-1})$. Upon receipt of $y_t$, and in the context of time-varying parameters, \cite{Katzfuss2019} \citep[see also][among others]{frei11} propose to use the observed-data likelihood under the EnKF to reweight parameter particles. We adapt the resulting particle ensemble Kalman filter (PEnKF) to our problem as follows. 

We impose the artificial evolution kernel (\ref{liu}) and consider a particle filter for recursive exploration of the target
\begin{align*}
&\widehat{\pi}_{\textrm{enkf},t}(\theta_{0:t},\tilde{x}_{0:t}^{1:N},x_{0:t}^{1:N}|y_{0:t}) \\
&\propto\pi(\theta_0)\widehat{p}_{\textrm{enkf},0}^{(\theta_0)}(y_0|\tilde{x}_0^{1:N})\left\{\prod_{s=1}^t \widehat{p}_{\textrm{enkf},s}^{(\theta_s)}(y_s|y_{0:s-1},\tilde{x}_s^{1:N})p_s(\theta_s|\theta_{s-1})\right\} \psi_t^{(\theta_{0:t})}(\tilde{x}_{0:t}^{1:N},x_{0:t}^{1:N})\\
&\propto \widehat{\pi}_{\textrm{enkf},t-1}(\theta_{0:t-1},\tilde{x}_{0:t-1}^{1:N},x_{0:t-1}^{1:N}|y_{0:t-1})\widehat{p}_{\textrm{enkf},t}^{(\theta_t)}(y_t|y_{0:t-1},\tilde{x}_t^{1:N})p_t(\theta_t|\theta_{t-1})\psi_t^{(\theta_{t})}(\tilde{x}_{t}^{1:N},x_{t}^{1:N}|x_{t-1}^{1:N})
\end{align*}
where $\psi_t^{(\theta_{t})}(\tilde{x}_{t}^{1:N},x_{t}^{1:N}|x_{t-1}^{1:N})$ is the joint density of the ensemble $\{\tilde{x}_{t}^{1:N},x_{t}^{1:N}\}$ (conditional on $\theta_{t}$ and $ x_{t-1}^{1:N}$) and can be sampled via application of Algorithm~\ref{alg:EnKF} with inputs $\theta_t$ and $x_{t-1}^{1:N}$. From a filtering perspective, marginalising $\widehat{\pi}_{\textrm{enkf},t-1}$ over $\theta_{0:t-2}$, $x_{0:t-2}^{1:N}$ and $\tilde{x}_{0:t-1}^{1:N}$ gives a particle filter that only requires storage of the weighted parameter particles $\{\theta_{t-1}^i,u_{t-1}^i\}$ (representing a weighted sample from the marginalised $\widehat{\pi}_{\textrm{enkf},t-1}$) and the associated ensemble members $x_{t-1}^{i,1:N}$, for $i=1,\ldots,M$, at time $t-1$. As with our implementation of SMC$^2$ in Section~\ref{subsec:smc2}, the storage cost of implementing PEnKF is then $\mathcal{O}(MN)$. 

The PEnKF assimilates the information in $y_t$ by first sampling  $\theta_t^{i}\sim p_t(\cdot|\theta_{t-1}^{i})$. The corresponding ensemble, $\{\tilde{x}_t^{i,1:N},x_t^{i,1:N}\}$, is drawn from 
$\psi_t^{(\theta_{t}^i)}(\cdot,\cdot|x_{t-1}^{i,1:N})$ by running 
Algorithm~\ref{alg:EnKF} with parameter $\theta_t^i$ and ensemble members $x_{t-1}^{i,1:N}$. Finally, a new weight is constructed  according to $u_t^i\propto u_{t-1}^i \widehat{p}_{\textrm{enkf},t}^{(\theta_t^i)}(y_t|y_{0:t-1},\tilde{x}_{t}^{i,1:N})$.

Step $t$ of the PEnKF is summarised by Algorithm~\ref{alg:PEnKF}. Note that we include a resampling step, which is triggered if the effective sample size (ESS) falls below some user specified threshold, $\gamma M$ . The ESS is calculated at time $t$ using (\ref{ess}).

\begin{algorithm}[htb] \caption{Particle ensemble Kalman filter (step $t$)} \label{alg:PEnKF}
\begin{algorithmic}
\STATE Input: weighted parameter particles $\{\theta_{t-1}^{i},u_{t-1}^{i}\}$ and associated ensemble members $x_{t-1}^{i,1:N}$, $i=1,\ldots,M$.
\FOR{$i=1,2,\ldots,M$}
\STATE \textbf{1. Update parameters}. Sample $\theta_t^{i}\sim \mathcal{N}\left(a\theta_{t-1}^{i}+(1-a)\bar{\theta}_{t-1}, h^2 V_{t-1}\right)$.
\STATE \textbf{2. Update weights}. Run Algorithm~\ref{alg:EnKF} with inputs $\theta_t^{i}$ and $x_{t-1}^{i,1:N}$ to obtain $\widehat{p}_{\textrm{enkf},t}^{(\theta_t^{i})}(y_t|y_{0:t-1},\tilde{x}_t^{i,1:N})$ and $x_{t}^{i,1:N}$. Construct the weight 
\[
\tilde{u}_{t}^{i}=u_{t-1}^{i}\widehat{p}_{\textrm{enkf},t}^{(\theta_t^{i})}(y_t|y_{0:t-1},\tilde{x}_t^{i,1:N}).
\]
\IF{$\textrm{ESS} < \gamma M$}
\STATE \textbf{3. Resample}. 
Normalise the weights to give $u_t^{i}=\tilde{u}_t^{i}/\sum_{k=1}^M \tilde{u}_t^{k}$. 
Sample an index $b_i$ from $\{1,\ldots,M\}$ with probabilities $u_t^{1:M}$. 
 Put $\{\theta_{t}^{i},u_t^{i}\}:=\{\theta_t^{b_i},1/M\}$ and $x_{t}^{i,1:N}:=x_{t}^{b_i,1:N}$.
\ENDIF
\ENDFOR
\STATE Output: weighted parameter particles 
$\{\theta_{t}^{i},u_{t}^{i}\}$ and associated ensemble members $x_{t}^{i,1:N}$, $i=1,\ldots,M$.
\end{algorithmic}
\end{algorithm}

\section{Nested ensemble Kalman filter (NEnKF)}
\label{sec:nest}
From the perspective of static parameter filtering, the EnKF-based methods described in Section~\ref{sec:KFinf} attempt to mitigate particle degeneracy via an artificial evolution kernel over the static parameters. This can be sensitive to the choice of tuning parameters \citep{vieira2016}. The SMC$^2$ algorithm described in Section~\ref{sec:bg} provides a more principled approach to alleviating particle degeneracy: if triggered at some time $t$, a resample-move step mutates parameter particles through a (pseudo-marginal) Metropolis-Hastings kernel that has $\pi_t(\theta|y_{0:t})$ as its invariant distribution. However, both the weighting and resample-move step in SMC$^2$ require running a particle filter over states, which can require a computationally prohibitive number of state particles $N$ in high-dimensional settings. For a specific example with $d_x=d_y=:n$, \cite{Katzfuss2019} show that the variance of the logarithm of observed-data likelihood under the EnKF, $\textrm{Var}[\log\widehat{p}_{\textrm{enkf},T}^{(\theta)}(y_{0:T}|\tilde{x}_{0:T}^{1:N})]$, can be controlled by scaling $N$ linearly with $n$, whereas $N$ should be scaled exponentially with $n$ when controlling $\textrm{Var}[\log\widehat{p}_{T}^{(\theta)}(y_{0:T}|x_{0:T}^{1:N},a_{0:T-1}^{1:N})]$, computed using a particle filter. When comparing the efficiency of ensemble MCMC and particle MCMC across several examples, \cite{drovandi22} find that the former can tolerate a much smaller value of $N$ compared to the latter. 

Our proposed approach leverages the robustness of the EnKF and the principled framework of SMC$^2$ by weighting static parameter particles by the EnKF likelihood and, where necessary, mutating these particles using the ensemble MCMC scheme described in Section~\ref{subsec:enkf}. By analogy with SMC$^2$, we term the resulting algorithm \emph{nested EnKF (NEnKF)}. In what follows, we provide the basic NEnKF algorithm before considering some extensions for improving efficiency.   

\subsection{NEnKF algorithm}\label{NEnKFbasic}
The NEnKF recursively targets $\widehat{\pi}_{\textrm{enkf},t}(\theta,x_{t}^{1:N}|y_{0:t})$ which can be constructed by taking the extended target
\begin{align*}
\widehat{\pi}_{\textrm{enkf},t}(\theta,\tilde{x}_{0:t}^{1:N},x_{0:t}^{1:N}|y_{0:t})
&\propto\pi(\theta)\widehat{p}_{\textrm{enkf},t}^{(\theta)}(y_{0:t}|\tilde{x}_{0:t}^{1:N}) \psi_t^{(\theta)}(\tilde{x}_{0:t}^{1:N},x_{0:t}^{1:N})\\
&\propto \widehat{\pi}_{\textrm{enkf},t-1}(\theta,\tilde{x}_{0:t-1}^{1:N},x_{0:t-1}^{1:N}|y_{0:t-1})\widehat{p}_{\textrm{enkf},t}^{(\theta)}(y_t|y_{0:t-1},\tilde{x}_t^{1:N})\\
&\times\psi_t^{(\theta)}(\tilde{x}_{t}^{1:N},x_{t}^{1:N}|x_{t-1}^{1:N})
\end{align*}
and marginalising out $\tilde{x}_{0:t}^{1:N}$ and $x_{0:t-1}^{1:N}$. Hence, given a weighted parameter sample $\{\theta^i,u_{t-1}^i\}$ with associated ensemble $x_{t-1}^{i,1:N}$, $i=1,\ldots,M$, from  $\widehat{\pi}_{\textrm{enkf},t-1}(\theta,x_{t-1}^{1:N}|y_{0:t-1})$, it should be clear that the weights are updated according to $u_t^i\propto u_{t-1}^i \widehat{p}_{\textrm{enkf},t}^{(\theta^i)}(y_t|y_{0:t-1},\tilde{x}_{t}^{i,1:N})$, by running Algorithm~\ref{alg:EnKF} with parameter $\theta^i$ and ensemble members $x_{t-1}^{i,1:N}$. 

Analagously to SMC$^2$, particle degeneracy is mitigated via a resample-move step. That is, when ESS (\ref{ess}) falls below a pre-specified threshold, $\gamma M$, we draw $M$ times from the mixture $\sum_{i=1}^M u_t^{i} {K}_{\textrm{enkf},t}\left(\theta^{i}\,,\, \cdot \right)$, 
where $\tilde{K}_t$ is an eMCMC kernel (see Section~\ref{subsec:eMCMC}) with target $\pi_{\text{enkf},t}(\theta|y_{0:t})$. Hence, after resampling, with $\theta^i$ denoting the $i$th resampled particle, a new particle $\theta^*$ is proposed from a kernel $q(\theta^{i},\cdot)$ and accepted with probability 
${\alpha}_{\textrm{enkf}}(\theta^*|\theta^{i}) = 1 \wedge \rho_{\textrm{enkf}}$, where
\begin{equation}\label{aprobEnKF}
  \rho_{\textrm{enkf}}:=
  \frac{\pi(\theta^*)}{\pi(\theta^{i})}\times \frac{\widehat{p}_{\textrm{enkf},t}^{(\theta^*)}(y_{0:t}|\tilde{x}_{0:t}^{*,1:N})}{\widehat{p}_{\textrm{enkf},t}^{(\theta^{i})}(y_{0:t}|\tilde{x}_{0:t}^{i,1:N})}\times    
\frac{q(\theta^*,\theta^{i})}{q(\theta^{i},\theta^*)}.
\end{equation}
Note that $\tilde{x}_{0:t}^{*,1:N}$ is implicitly generated by running the EnKF forward from time $0$. In what follows, we take $q(\theta^{i},\cdot)$ to be the kernel associated with a random walk Metropolis move of the form
\[
\theta^*=\theta^{i}+\epsilon^{i}, \qquad \epsilon^{i}\sim \mathcal{N}(0,\zeta V_t).
\]
To ensure a reversible move, the variance $V_t$ can be computed as the sample variance of all resampled particles except the $i$th. The scaling $\zeta$ can be chosen by the practitioner or set using a rule of thumb, for example, in the pseudo-marginal context, \cite{sherlock2015} suggest $\zeta^2=2.56^2/d_{\theta}$. Efficiency of the move step can be increased, at increased computational cost, by repeatedly sampling from ${K}_{\textrm{enkf},t}\left(\theta^{i}\,,\, \cdot \right)$. Although we do not pursue it here, adapting the number of MCMC iterations in the mutation step is discussed in the SMC$^2$ context in \cite{botha23}. Step $t$ of the NEnKF can be found in Algorithm~\ref{alg:NEnKF}. Note that for ease of exposition, we present the algorithm for a fixed value $N$ and for a single iteration of eMCMC (per parameter particle) in the mutation step. 

\begin{algorithm}[htb] \caption{Nested ensemble Kalman filter (step $t$)} \label{alg:NEnKF}
\begin{algorithmic}
\STATE Input: weighted parameter particles $\{\theta^{i},u_{t-1}^{i}\}$, associated ensemble members $x_{t-1}^{i,1:N}$ and likelihood evaluations $\widehat{p}_{\textrm{enkf},t-1}^{(\theta^{i})}(y_{0:t-1}|\tilde{x}_{0:t-1}^{i,1:N})$, $i=1,\ldots,M$.
\FOR{$i=1,2,\ldots,M$}
\STATE \textbf{1. Update weights and likelihood}. Run Algorithm~\ref{alg:EnKF} with inputs $\theta^{i}$ and $x_{t-1}^{i,1:N}$ to obtain $\widehat{p}_{\textrm{enkf},t}^{(\theta^{i})}(y_t|y_{0:t-1},\tilde{x}_t^{i,1:N})$ and $x_{t}^{i,1:N}$. Construct the weight 
\[
\tilde{u}_{t}^{i}=u_{t-1}^{i}\widehat{p}_{\textrm{enkf},t}^{(\theta^{i})}(y_t|y_{0:t-1},\tilde{x}_{t}^{i,1:N})
\]
and update the likelihood 
\[
\widehat{p}_{\textrm{enkf},t}^{(\theta^{i})}(y_{0:t}|\tilde{x}_{0:t}^{i,1:N})=\widehat{p}_{\textrm{enkf},t-1}^{(\theta^{i})}(y_{0:t-1}|\tilde{x}_{0:t-1}^{i,1:N})\widehat{p}_{\textrm{enkf},t}^{(\theta^{i})}(y_t|y_{0:t-1},\tilde{x}_t^{i,1:N}).
\]
\IF{$\textrm{ESS} < \gamma M$}
\STATE \textbf{2. Resample}. 
Normalise the weights to give $u_t^{i}=\tilde{u}_t^{i}/\sum_{k=1}^M \tilde{u}_t^{k}$. 
 Sample an index $b_i$ from $\{1,\ldots,M\}$ with probabilities $u_t^{1:M}$. 
 Put $\{\theta^{i},u_t^{i}\}:=\{\theta^{b_i},1/M\}$ and $x_{t}^{i,1:N}:=x_{t}^{b_i,1:N}$.
 \STATE \textbf{3. Move / mutate}. Propose $\theta^*\sim q(\theta^{i},\cdot)$. Perform iterations $1,\ldots,t$ of Algorithm~\ref{alg:EnKF} to obtain $\widehat{p}_{\textrm{enkf},t}^{(\theta^{*})}(y_{0:t}|\tilde{x}_{0:t}^{*,1:N})$ and $x_{t}^{*,1:N}$. With probability ${\alpha}_{\textrm{enkf}}(\theta^*|\theta^{i}) = 1 \wedge \rho_{\textrm{enkf}}$, where $\rho_{\textrm{enkf}}$ is given by (\ref{aprobEnKF}), put $\theta^{i}=\theta^*$, $\widehat{p}_{\textrm{enkf},t}^{(\theta^{i})}(y_{0:t}|\tilde{x}_{0:t}^{i,1:N})=\widehat{p}_{\textrm{enkf},t}^{(\theta^{*})}(y_{0:t}|\tilde{x}_{0:t}^{*,1:N})$ and $x_{t}^{i,1:N}=x_{t}^{*,1:N}$.
\ENDIF
\ENDFOR
\STATE Output: weighted parameter particles 
$\{\theta^{i},u_{t}^{i}\}$, associated ensemble members $x_{t}^{i,1:N}$ and likelihood evaluations $\widehat{p}_{\textrm{enkf},t}^{(\theta^{i})}(y_{0:t}|\tilde{x}_{0:t}^{i,1:N})$, $i=1,\ldots,M$.
\end{algorithmic}
\end{algorithm}

Following the approach used in SMC$^2$, we also consider a mechanism for dynamically updating the number of ensemble members $N$. After completion of the mutation step, if updating $N$ is triggered then for each particle $\theta^i$, ensemble members $x_{0:t}^{i,1:N}$ are replaced by $\check{x}_{0:t}^{i,1:\check{N}}$ using the generalised importance sampling strategy of \cite{delMoral06}, with an artificial backward kernel chosen to give an incremental weight of 1 \citep{botha23}. Hence, for each particle $\theta^i$, a new ensemble $\check{x}_{0:t}^{i,1:\check{N}}$ is generated through repeated application of Algorithm~\ref{alg:EnKF}. We adapt the method of \cite{duan15} to our context, 
by updating $N$ if $\hat{\sigma}_N^2>1.5$, where $\hat{\sigma}_N^2$ is an estimate of the variance of $\log\widehat{p}_{\text{enkf},t}^{(\bar{\theta})}(y_{0:t}|\tilde{x}_{0:t}^{1:N})$, based on, say, $r$ runs of the EnKF with $\theta$ fixed at an estimate of some central posterior value (e.g. the posterior mean). When the updating of $N$ is triggered, we set  $\check{N}=\hat{\sigma}_N^2 N$. We refer the reader to \cite{botha23} for further discussion, noting their critique that, in the SMC$^2$ setting, this approach can be sensitive to the initial value of $N$. However, and as discussed at the beginning of this section, $\widehat{p}_{\text{enkf},t}^{(\theta)}(y_{0:t}|\tilde{x}_{0:t}^{1:N})$ is likely to be less sensitive to $N$ than the corresponding particle filter estimator. 

\subsection{Resample-move: delayed-acceptance kernel}\label{da}
As with SMC$^2$, the most computationally expensive step of the NEnKF algorithm is the resample-move step; if executed at time $t$, the cost is $\mathcal{O}(tMN)$. We, therefore, aim to avoid expensive calculation of the likelihood $\widehat{p}_{\text{enkf},t}^{(\theta)}(y_{0:t}|\tilde{x}_{0:t}^{1:N})$ for proposed parameter values that are likely to be rejected. Hence, we propose to replace the standard Metropolis-Hastings move in step 3 of Algorithm~\ref{alg:NEnKF} with a delayed acceptance (DA) move \citep[\emph{e.g.}][]{christen2005,Goli15,Banterle19}. 

The DA move requires a surrogate for which we use the (unique) resampled particles and their corresponding log-likelihood evaluations. Given $\{\theta^{i},\log \widehat{p}_{\text{enkf}}^{(\theta^{i})}(y_{0:t}|\tilde{x}_{0:t}^{i,1:N})\}_{i=1}^{M_r}$, where $M_r$ denotes the number of unique resampled parameter particles, we construct an approximate likelihood for a new parameter value, $\theta$, as follows. We find the $k$-nearest neighbours of $\theta$, denoted $\theta^{[1]},\ldots,\theta^{[k]}$, and their associated log-likelihood estimates $\log \widehat{p}_{\text{enkf}}^{(\theta^{[1]})}(y_{0:t}|\tilde{x}_{0:t}^{[1],1:N}),\ldots,\log\widehat{p}_{\text{enkf}}^{(\theta^{[k]})}(y_{0:t}|\tilde{x}_{0:t}^{[k],1:N})$. We then approximate the log-likelihood via the weighted average
\[
\log \widehat{p}_{\textrm{knn},t}^{(\theta)}(y_{0:t})=\frac{\sum_{i=1}^k d_{[i]}^{-1}\log \widehat{p}_{\text{enkf},t}^{(\theta^{[i]})}(y_{0:t}|\tilde{x}_{0:t}^{[i],1:N})}{\sum_{i=1}^k d_{[i]}^{-1}}
\]
where $d_{[i]}$ denotes the distance (using some metric) between $\theta$ and $\theta^{[i]}$. Note that we suppress explicit dependence of the surrogate on the $k$-nearest neighbours for notational simplicity.

For a given particle $\theta^i$, Stage One of the DA scheme proposes $\theta^* \sim q(\theta^i,\cdot)$, computes $\widehat{p}_{\textrm{knn},t}^{(\theta^*)}(y_{0:t})$ and the screening acceptance probability $\alpha_1\left(\theta^i,\theta^*\right) = 1 \wedge \rho_1$, where
\begin{equation} \label{stage1}
 \rho_1:=\frac{\pi(\theta^*) \widehat{p}_{\textrm{knn},t}^{(\theta^*)}(y_{0:t})}{\pi(\theta^i) \widehat{p}_{\textrm{knn},t}^{(\theta^i)}(y_{0:t})}\times \frac{q(\theta^*,\theta^i)}{q(\theta^i,\theta^*)} .
\end{equation}
If this screening step is successful, Stage Two of the DA scheme constructs $\widehat{p}_{\text{enkf},t}^{(\theta^*)}(y_{0:t}|\tilde{x}_{0:t}^{*,1:N})$ and the Stage Two acceptance probability $\alpha_{2|1}(\theta^i,\theta^*) 
= 1 \wedge \rho_{2|1}$, where 
\begin{equation} \label{stage2}
\rho_{2|1}:=\frac{\widehat{p}_{\text{enkf},t}^{(\theta^*)}(y_{0:t}|\tilde{x}_{0:t}^{*,1:N})}{\widehat{p}_{\text{enkf},t}^{(\theta^i)}(y_{0:t}|\tilde{x}_{0:t}^{i,1:N})}\times \frac{\widehat{p}_{\textrm{knn},t}^{(\theta^i)}(y_{0:t})}{ \widehat{p}_{\textrm{knn},t}^{(\theta^*)}(y_{0:t})} .
\end{equation}
The overall acceptance probability for the scheme is $\alpha_1\alpha_{2|1}$ 
and standard arguments show that iterating the DA scheme defines a Markov chain that is reversible with respect to the (extended) target $\widehat{\pi}_{\textrm{enkf},t}(\theta,\tilde{x}_{0:t}^{1:N},x_{0:t}^{1:N}|y_{0:t})$.

By design (to ensure detailed balance), $\rho_{1}\rho_{2|1}=\rho_{\textrm{enkf}}$. Hence $\alpha_1\alpha_{2|1}\le \alpha_{\textrm{enkf}}$, since for any $a,b\ge 0$, $\{1\wedge a\}\{1\wedge b\}\le 1\wedge ab$. Thus, although the delayed acceptance step can substantially reduce the number of computationally intensive and potentially wasteful estimations of the EnKF likelihood, it also reduces the expected amount of movement. Intuitively, this reduction will be small when $\rho_{2|1}\approx 1$. More formally, we may obtain a bound on the worst possible behaviour:
\[
\alpha_{1}
=
1\wedge \left\{\rho_{\textrm{enkf}}\times \frac{1}{\rho_{2|1}}\right\}
\ge
\left\{1\wedge \rho_{\textrm{enkf}}\right\}\times \left\{1\wedge\frac{1}{\rho_{2|1}}\right\}
=
\alpha_{\textrm{enkf}}\times \left\{1\wedge\frac{1}{\rho_{2|1}}\right\}.
\]
This provides a lower bound for $\alpha_{\textrm{da}}$, the overall acceptance probability of the delayed acceptance move: 
\[
\alpha_{\textrm{da}}=\alpha_1\alpha_{2|1}
\ge
\alpha_{\textrm{enkf}}\times \left\{1\wedge\frac{1}{\rho_{2|1}}\right\}\times \left\{1\wedge \rho_{2|1}\right\}
=
\alpha_{\textrm{enkf}}\times\exp\{-|\log \rho_{2|1}|\}.
\]
In general, $|\log\rho_{2|1}|$ is small when
\[
\widehat{p}_{\textrm{knn},t}^{(\theta^i)}(y_{0:t})
  \approx
  \widehat{p}_{\text{enkf},t}^{(\theta^i)}(y_{0:t}|\tilde{x}_{0:t}^{i,1:N})
~~~\mbox{and}~~~
\widehat{p}_{\textrm{knn},t}^{(\theta^*)}(y_{0:t})
\approx
\widehat{p}_{\text{enkf},t}^{(\theta^*)}(y_{0:t}|\tilde{x}_{0:t}^{*,1:N}).
\]
With our inverse-distance weighting scheme, the first approximate equality is, in fact, an equality. A necessary condition for the second approximate equality is that both $\widehat{p}_{\text{enkf},t}^{(\theta^*)}(y_{0:t}|\tilde{x}_{0:t}^{*,1:N})$ and $\widehat{p}_{\textrm{knn},t}^{(\theta^*)}(y_{0:t})$ are close to $E^{\theta^*}_t:=\text{E}\left[\widehat{p}_{\text{enkf},t}^{(\theta^*)}(y_{0:t}|\tilde{X}_{0:t}^{*,1:N})\right]$. Keeping $\hat{\sigma}_N^2<1.5$ helps ensure closeness for the former, most of the time; for example, if $\widehat{p}_{\text{enkf},t}^{(\theta^*)}(y_{0:t}|\tilde{X}_{0:t}^{*,1:N})$ has a lognormal distribution then $\mathbb{P}\left(|\log \widehat{p}_{\text{enkf},t}^{(\theta^*)}(y_{0:t}|\tilde{x}_{0:t}^{*,1:N})-\log E^{\theta^*}_t|<1\right)>1/2$. The $k$-nearest neighbour approximation averages $k$ noise terms, so the variability in this approximation to $\log E^{\theta^*}_t$ will typically be substantially less than $1.5$. Thus, we expect the discrepancy here to be driven by how well the weighted average of $E^{\theta^{[i]}}_t,\dots,E^{\theta^{[k]}}_t$ approximates $E^{\theta^*}_t$. This will hold provided $\log E^{\theta}_t$ varies sufficiently slowly with $\theta$ compared with the spacing of the $N$ points. Thus, we would expect the $k$-nearest neighbour strategy to fail when the parameter space is high-dimensional or when the log posterior experiences high-amplitude oscillations over short distances.

\subsection{Nonlinear observation model}\label{NGobs}

It has been assumed that the density, $f_t^{(\theta)}(\cdot|x)$, arising from the observation model is a linear and Gaussian conditional density; see (\ref{eq:gauss}). Suppose now that ${f}_t^{(\theta)}$ can be evaluated point-wise and may be neither linear nor Gaussian. Key to our approach in this scenario is that ${f}_t^{(\theta)}$ can be approximated by a linear Gaussian conditional density which we denote by $g_t^{(\theta)}(\cdot|x)$ and whose form is given by (\ref{eq:gauss}). That is, we take
\[
g_t^{(\theta)}(y_t|x_t)= \mathcal{N}(y_t; \tilde{H} x_t, \tilde{R})
\]
for suitable choice of $\tilde{H}$ and $\tilde{R}$. For example, if ${f}_t^{(\theta)}(y|x)=\mathcal{N}(y_t; h(x_t), R)$ for some nonlinear function $h(\cdot)$ (which may depend on $\theta$), an approximate density $g_t^{(\theta)}$ can be found be taking the first two terms of a Taylor series expansion of $h(\cdot)$ about the forecast mean $\mu_{t|t}$ \citep[see e.g.][]{evensen2022}. In this case, $\tilde{H}$ is a Jacobian matrix and we also include the offset term $h(\mu_{t|t})-\tilde{H}\mu_{t|t}$ in the observation mean. Alternatively, $x_t$ can be augmented to include $h(\cdot)$ in the forecast ensemble \citep[see e.g.][]{anderson01}. Then, $x_t$ is replaced by $z_t=(x_t',h(x_t)')'$ and $\tilde{H}=(0_{d_x\times d_x}, I)$ for an identity matrix $I$ of appropriate dimension. Examples of appropriate $g_t^{(\theta)}$ in the case of a Gaussian observation model with state-dependent variance can be found in Sections~\ref{sec:app2} and \ref{sec:app2b}.      

We consider first the state estimation problem within the context described above. We denote by $\widetilde{p}_{\textrm{enkf},t}^{(\theta)}(x_t|y_{0:t},\tilde{x}_{t}^{1:N})$ the filtering density at time $t$, based on use of the EnKF with a \emph{nonlinear observation model} ${f}_t^{(\theta)}$. We obtain
\begin{equation}\label{eq:filtNL}
\widetilde{p}_{\textrm{enkf},t}^{(\theta)}(x_t|y_{0:t},\tilde{x}_{t}^{1:N})= \frac{\widehat{p}_{\textrm{enkf},t}^{(\theta)}(x_t|y_{0:t-1},\tilde{x}_{t}^{1:N}){f}_t^{(\theta)}(y_t|x_t)}{\widetilde{p}_{\textrm{enkf},t}^{(\theta)}(y_t|y_{0:t-1},\tilde{x}_t^{1:N})}
\end{equation}
which has the same form as (\ref{eq:filt}), but now owing to the nonlinear observation model, the denominator (that is, the observed-data likelihood contribution)  $\widetilde{p}_{\textrm{enkf},t}^{(\theta)}(y_t|y_{0:t-1},\tilde{x}_t^{1:N})$ is intractable. The filtering density in (\ref{eq:filtNL}) can be approximated by replacing ${f}_t^{(\theta)}$ with the linear and Gaussian approximation ${g}_t^{(\theta)}$. The resulting approximation is a tractable Gaussian density given by
\begin{equation}\label{eq:filtNLg}
\widehat{g}_{\textrm{enkf},t}^{(\theta)}(x_t|y_{0:t},\tilde{x}_{t}^{1:N})= \frac{\widehat{p}_{\textrm{enkf},t}^{(\theta)}(x_t|y_{0:t-1},\tilde{x}_{t}^{1:N}){g}_t^{(\theta)}(y_t|x_t)}{\widehat{g}_{\textrm{enkf},t}^{(\theta)}(y_t|y_{0:t-1},\tilde{x}_t^{1:N})}.
\end{equation}
Hence, the denominator $\widehat{g}_{\textrm{enkf},t}^{(\theta)}(y_t|y_{0:t-1},\tilde{x}_t^{1:N})$ can be computed analytically using Algorithm~\ref{alg:EnKF}, with $H=\tilde{H}$ and $R=\tilde{R}$, deduced from the example specific ${g}_t^{(\theta)}$. Parameter (and state) filtering may then proceed via the NEnKF (Algorithm~\ref{alg:NEnKF}) at the expense of inferential accuracy, due to the additional error induced by ${g}_t^{(\theta)}$. 

Naturally, inferential accuracy can be maintained at the expense of computational cost, by eschewing the  EnKF in favour of the particle filter with target ${p}_t^{(\theta)}(x_{t}|y_{0:t})$ given by (\ref{filt}). That is, Algorithm~\ref{alg:PartF} can be run, for example, with a proposal density $q_t^{(\theta)}(x_t|x_{t-1},y_t):=p_t(x_t|x_{t-1})$. However, particles propagated myopically of the next observation can result in an observed-data likelihood estimator with high variance \citep[][]{delmoral15}, which can be inefficient when used inside SMC$^2$ \citep{golikyp18}. Methods which use the EnKF as a proposal mechanism inside a particle filter over states have been proposed by \cite{papadakis2010} \citep[see also][]{tamboli2021,leeuwen2019}. In particular, use of $\widehat{g}_{\textrm{enkf},t}^{(\theta)}(x_t|y_{0:t},\tilde{x}_{t}^{1:N})$ in place of $q_t^{(\theta)}(x_t|x_{t-1}^{a_{t-1}^j},y_t)$ in Algorithm~\ref{alg:PartF} corresponds to the weighted ensemble Kalman filter of \cite{papadakis2010}. One possible implementation of this approach is to take $\widehat{g}_{\textrm{enkf},t}^{(\theta)}(x_t|y_{0:t},\tilde{x}_{t}^{1:N})$ as the density associated with an ensemble member generated by the update step of Algorithm~\ref{alg:EnKF}, with $H=\tilde{H}$ and $R=\tilde{R}$, and conditional on the forecast ensemble $\tilde{x}_{t}^{1:N}$. The weight function then takes the form
\begin{align}\label{weight0}
\tilde{w}_{t}(\tilde{x}_{t}^{1:N},x_{t-1}^{a_{t-1}^j},x_t^{j})=\frac{p_t^{(\theta)}(x_t^{j}|x_{t-1}^{a_{t-1}^j})f_t^{(\theta)}(y_t|x_t^{j})}{\widehat{g}_{\textrm{enkf},t}^{(\theta)}(x_t^j|y_{0:t},\tilde{x}_{t}^{1:N})}.
\end{align}
However, in many scenarios of practical interest, it is beneficial to model states at a finer frequency than at which observations are observed. In this case, the state vector $x_t$ is replaced by the matrix $(x_{t,1},\ldots,x_{t,m})$, whose elements can be simulated recursively from some kernel $p_{t,i}^{(\theta)}(x_{t,i}|x_{t,i-1})$ such that $p_t^{(\theta)}(x_t|x_{t-1})=\int \prod_{i=1}^m p_{t,i}^{(\theta)}(x_{t,i}|x_{t,i-1})dx_{t,1:m-1}$. In this setting, $p_t^{(\theta)}(x_t|x_{t-1})$ can be sampled but not typically evaluated, precluding the use of (\ref{weight0}).

We therefore propose a third option, which leverages the inferential accuracy of SMC$^2$ and the computational efficiency of the EnKF, and can be applied in scenarios where the latent state process is required on a finer time-scale than the observations. Here, the inner loop of SMC$^2$ operates by using a particle filter over states, with target $\widetilde{p}_{\textrm{enkf},t}^{(\theta)}(x_t|y_{0:t},\tilde{x}_{t}^{1:N})$ given by (\ref{eq:filtNL}) and a proposal mechanism $\widehat{g}_{\textrm{enkf},t}^{(\theta)}(x_t|y_{0:t},\tilde{x}_{t}^{1:N})$ given by (\ref{eq:filtNLg}). Hence, in Algorithm~\ref{alg:PartF}, the propagate step samples $x_t^j\sim \widehat{g}_{\textrm{enkf},t}^{(\theta)}(x_t|y_{0:t},\tilde{x}_{t}^{1:N})$, $j=1\ldots,N$, via a single run of Algorithm~\ref{alg:EnKF} with inputs $\theta$, resampled particles $x_{t-1}^{a_{t-1}^{1:N}}$, $H=\tilde{H}$ and $R=\tilde{R}$. It should be clear that the unnormalised weight of the $j$th particle $x_t^j$ is given by
\begin{align}\label{weight}
\tilde{w}_{t}(\tilde{x}_{t}^{1:N},x_t^{j})&=\frac{\widehat{p}_{\textrm{enkf},t}^{(\theta)}(x_t^j|y_{0:t-1},\tilde{x}_{t}^{1:N}){f}_t^{(\theta)}(y_t|x_t^j)}
{\widehat{g}_{\textrm{enkf},t}^{(\theta)}(x_t^j|y_{0:t},\tilde{x}_{t}^{1:N})} \nonumber\\
&= \widehat{g}_{\textrm{enkf},t}^{(\theta)}(y_t|y_{0:t-1},\tilde{x}_t^{1:N}) \times \frac{{f}_t^{(\theta)}(y_t|x_t^j)}
{{g}_t^{(\theta)}(y_t|x_t^j)}
\end{align}
where $\widehat{g}_{\textrm{enkf},t}^{(\theta)}(y_t|y_{0:t-1},\tilde{x}_t^{1:N})$ is obtained from the single run of Algorithm~\ref{alg:EnKF} in the propagate step. We note that $p_t^{(\theta)}(x_t|x_{t-1})$ is sampled in step 1 of Algorithm~\ref{alg:EnKF} when generating the forecast ensemble, but need never be evaluated, since the corresponding forecast density, $\widehat{p}_{\textrm{enkf},t}^{(\theta)}(x_t|y_{0:t-1},\tilde{x}_{t}^{1:N})$, appears in both $\widetilde{p}_{\textrm{enkf},t}^{(\theta)}(x_t|y_{0:t},\tilde{x}_{t}^{1:N})$ and $\widehat{g}_{\textrm{enkf},t}^{(\theta)}(x_t|y_{0:t},\tilde{x}_{t}^{1:N})$, and therefore cancels in the weight function. This necessarily sacrifices inferential accuracy (compared to targeting ${p}_t^{(\theta)}(x_{t}|y_{0:t})$) but allows for integrating $x_{t-1}$ (and $(x_{t,1},\ldots,x_{t,m})$ where relevant) out of the weight function. This procedure is known as Rao-Blackwellisation (RB) and can lead to weights with lower variance \citep[see e.g.][]{doucet2000}. Step $t$ of the Rao-Blackwellised particle filter is given by Algorithm~\ref{alg:PartRB}. 
\begin{algorithm}[htb] \caption{Rao-Blackwellised particle filter (step $t$)} \label{alg:PartRB}
\begin{algorithmic}
\STATE Input: parameter $\theta$ and weighted particles $\{x_{t-1}^{1:N},w_{t-1}^{1:N}\}$.
\STATE \textbf{1. Resample}. For $j=1,2,\ldots,N$, sample an index $a_{t-1}^j$ from $\{1,\ldots,N\}$ with probabilities $w_{t-1}^{1:N}$ and put $x_{t-1}^j:=x_{t-1}^{a_{t-1}^j}$. 
\STATE \textbf{2. Propagate}. Run Algorithm~\ref{alg:EnKF} with inputs $\theta$, ensemble members $x_{t-1}^{1:N}$, $H=\tilde{H}$ and $R=\tilde{R}$ to obtain ${x}_t^{1:N}\sim \widehat{g}_{\textrm{enkf},t}^{(\theta)}(x_t|y_{0:t},\tilde{x}_{t}^{1:N})$ and $\widehat{g}_{\textrm{enkf},t}^{(\theta)}(y_t|y_{0:t-1},\tilde{x}_t^{1:N})$.
\STATE \textbf{3. Weight}. For $j=1,2,\ldots,N$, construct and normalise the weight 
\[
\tilde{w}_{t}(\tilde{x}_{t}^{1:N},x_t^{j})= \widehat{g}_{\textrm{enkf},t}^{(\theta)}(y_t|y_{0:t-1},\tilde{x}_t^{1:N}) \times \frac{{f}_t^{(\theta)}(y_t|x_t^j)}
{{g}_t^{(\theta)}(y_t|x_t^j)},
 \quad w_t^{j}=\tilde{w}_{t}(\tilde{x}_{t}^{1:N},x_t^{j})/\sum_{k=1}^{N}\tilde{w}_{t}(\tilde{x}_{t}^{1:N},x_t^{k})
.\] 
\STATE Output: likelihood estimate $\widehat{\widetilde{p}}_{\textrm{enkf},t}^{(\theta)}(y_t|y_{0:t-1},\tilde{x}_t^{1:N}) = \frac{1}{N}\sum_{j=1}^N \tilde{w}_{t}(x_{t}^{1:N},x_t^{j})$ and weighted particles $\{x_{t}^{1:N},w_t^{1:N}\}$.
\end{algorithmic}
\end{algorithm}

Finally, taking the average (over state particles $x_t^{1:N}$) of $\tilde{w}_{t}(\tilde{x}_{t}^{1:N},x_t^{j})$ in (\ref{weight}) estimates the observed-data likelihood $\widetilde{p}_{\textrm{enkf},t}^{(\theta)}(y_t|y_{0:t-1},\tilde{x}_t^{1:N})$, that is, the denominator in (\ref{eq:filtNL}). Such estimates can be used directly in the SMC$^2$ scheme (see Algorithm~\ref{alg:smc2}) to give recursive exploration of a target which we denote by $\widetilde{\pi}_{\textrm{enkf},t}(\theta|y_{0:t})$. We refer to the resulting inference scheme for this target as RB-SMC$^2$. Given that the inner particle filter of RB-SMC$^2$ corrects for the use of $g_{t}^{(\theta)}$ via reweighting, we anticipate better inferential accuracy than NEnKF (which uses $g_{t}^{(\theta)}$ without further correction), albeit at increased computational cost. Provided that a suitable $g_{t}^{(\theta)}$ can be sought, we expect that RB-SMC$^2$ will provide greater computational efficiency than SMC$^2$, albeit for a reduction in inferential accuracy. It therefore represents a compromise between NEnKF and SMC$^2$. We investigate the performance of RB-SMC$^2$ in Sections~\ref{sec:app2} and \ref{sec:app2b}.    

\section{Applications}
\label{sec:app}

To illustrate the nested ensemble Kalman filter (NEnKF) and benchmark against existing methods, we consider four applications of increasing complexity. In Section~\ref{sec:app1}, we perform inference on an Ornstein-Uhlenbeck model using synthetic data. We compare the performance of NEnKF against the augmented and particle  ensemble Kalman filters (AEnKF and PEnKF of Sections~\ref{subsec:aug} and \ref{subsec:part}). We additionally benchmark performance and accuracy using the output of an MCMC scheme which directly targets the parameter posterior. In Section~\ref{sec:app2}, we apply the NEnKF and two implementations of SMC$^2$ in the context of a nonlinear observation model, via a stochastic model of predator-prey interaction. Section~\ref{sec:app2b} considers a real data example involving the spread of oak processionary moth in south-east England. Finally, in Section~\ref{sec:app3}, we consider inference for the parameters governing a 10-dimensional Lorenz-96 model. For the second and final applications, inferential accuracy is benchmarked against the output of a particle MCMC scheme. All algorithms were coded in R and run on a desktop computer with an Intel Core i7-10700 processor and a 2.90GHz clock speed. For the Lorenz-96 application (Section~\ref{sec:app3}), the per particle steps corresponding to reweighting, mutation and updating of $N$ were parallelised across 10 cores.  The code is available from https://github.com/AndyGolightly/NestedEnKF.

\subsection{Ornstein-Uhlenbeck process}\label{sec:app1}

Consider a process $\{X_t,t \geq 0\}$ satisfying an It\^o stochastic differential equation (SDE) of the form
\[
dX_t = \theta_1 (\theta_2-X_t)dt + \theta_3 dW_t
\]
where $W_t$ is a standard Brownian motion process. This SDE can be solved analytically to give 
\[
p(x_{t+\Delta t}|x_t,\theta)=\mathcal{N}\left(x_{t+\Delta t};\, x_t e^{-\theta_1 \Delta t} +\theta_2 (1-e^{-\theta_1 \Delta t}) \,,\, \frac{\theta_3^2}{2\theta_1}(1-e^{-2\theta_1 \Delta t})\right).
\]
Using this solution, we simulated 50 points at integer times from a single realisation of a path using an initial condition of $x_0=10$ and $\theta=(1,2,1)'$. We then corrupted the resulting skeleton path using the observation model in (\ref{eq:gauss}) with $H=1$ and $R=0.1$. We adopted an independent prior specification with $\theta_1\sim \text{Gamma}(2,2)$, $\theta_2\sim \text{Gamma}(5,3)$ and $\theta_3\sim \text{Gamma}(2,5)$. In what follows, we assume that the initial condition and observation noise are fixed and known. 

We ran NEnKF with $M=1000$ particles and initial ensemble members $N=10$. A mutation step was triggered if the effective sample size (\ref{ess}) fell below 400 ($\gamma=0.4$). Updating of $N$ was triggered following a move step if the estimate $\hat{\sigma}_N^2$ of $\text{Var}[\log\widehat{p}_{\text{enkf},t}^{(\bar{\theta})}(y_{0:t}|\tilde{x}_{0:t}^{1:N})]$ exceeded 1.5, resulting in the update $\tilde{N}=\hat{\sigma}_N^2 N$. We additionally ran NEnKF with a delayed acceptance kernel in the move step, based on the $k$-nearest neighbours surrogate of Section~\ref{da}. We used $k=3$, which gave a good balance between computational efficiency and accuracy of the surrogate. This gave a modest decrease in CPU time of around 10\% over NEnKF; therefore all results are based on NEnKF with delayed acceptance.     

We ran PEnKF and AEnKF with respective particle $M$ and ensemble member $N$ values chosen to give a CPU time in approximate agreement with that of NEnKF-DA. For PEnKF, we fixed $N$ at the average value obtained at termination of NEnKF over several replicate runs. Both schemes require choosing the hyper-parameters $a$ and $h$ used in the artificial evolution kernel. We followed \cite{liu2001} by setting $a=\sqrt{1-h^2}$ and $h=1-((3\delta-1)/2\delta)^2$ with the recommendation that the discount factor $\delta$ is around $0.95-0.99$. We considered several choices of $\delta$ and report results for $\delta=0.97$, which we found to give best performance.



\begin{table}[t]
\centering
\small
	\begin{tabular}{@{}l llll lllll@{}}
         \toprule
Scheme  & $N$ & $M$ &  CPU & \multicolumn{6}{l}{Bias (RMSE)}  \\
\cmidrule(l){5-10}
	& &         &       &$\widehat{\text{E}}(\log\theta_1)$     & $\widehat{\text{E}}(\log\theta_2)$ & $\widehat{\text{E}}(\log\theta_3)$ &$\widehat{\text{SD}}(\log\theta_1)$  &$\widehat{\text{SD}}(\log\theta_2)$ & $\widehat{\text{SD}}(\log\theta_3)$\\   
\midrule
NEnKF    & 100\phantom{00} & 1000  &43 &\phantom{-}0.0036  &-0.0047& \phantom{-}0.0003&0.0068 &0.0014& \phantom{-}0.0003\\
 & & & & \phantom{-}(0.031) & \phantom{-}(0.010)& \phantom{-}(0.021) & (0.019)  &(0.005)&\phantom{-}(0.010)\\
PEnKF    & 100\phantom{00} & 4000  &46 &\phantom{-}0.0027&-0.0078& \phantom{-}0.0037&0.0098&0.0010& -0.0020 \\
& & & & \phantom{-}(0.033) & \phantom{-}(0.010)& \phantom{-}(0.024) & (0.020)  &(0.007)&\phantom{-}(0.012)\\
AEnKF    & 25000   & \phantom{11}--&50 &-0.0861 &-0.0396&-0.8898&0.0243 &0.0058& \phantom{-}0.6795  \\
& & & & \phantom{-}(0.020) & \phantom{-}(0.006)& \phantom{-}(0.068) & (0.004)  &(0.004)&\phantom{-}(0.005)\\
\bottomrule
\end{tabular}
	\caption{OU example. Number of ensemble members $N$ (at termination), number of particles $M$, CPU time (in seconds), bias (and RMSE in parentheses) of estimators 
of the posterior expectations $\text{E}(\log\theta_i)$ and standard deviations $\text{SD}(\log\theta_i)$, $i=1,2,3$. All results are obtained by averaging over 100 runs of each inference scheme.}\label{tab:tabOU}
\end{table}

\begin{figure}[ht]
\centering
\includegraphics[width=15cm,height=4.0cm]{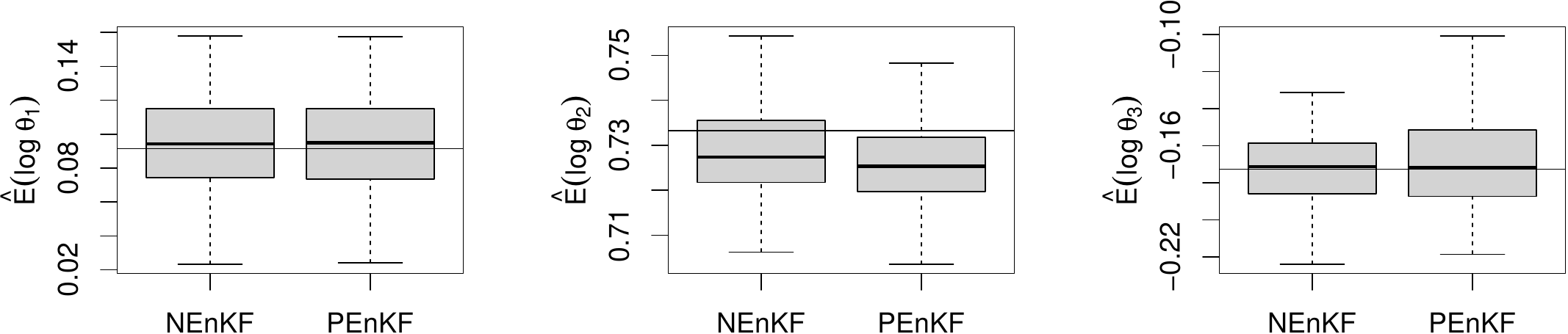}

\

\includegraphics[width=15cm,height=3.7cm]{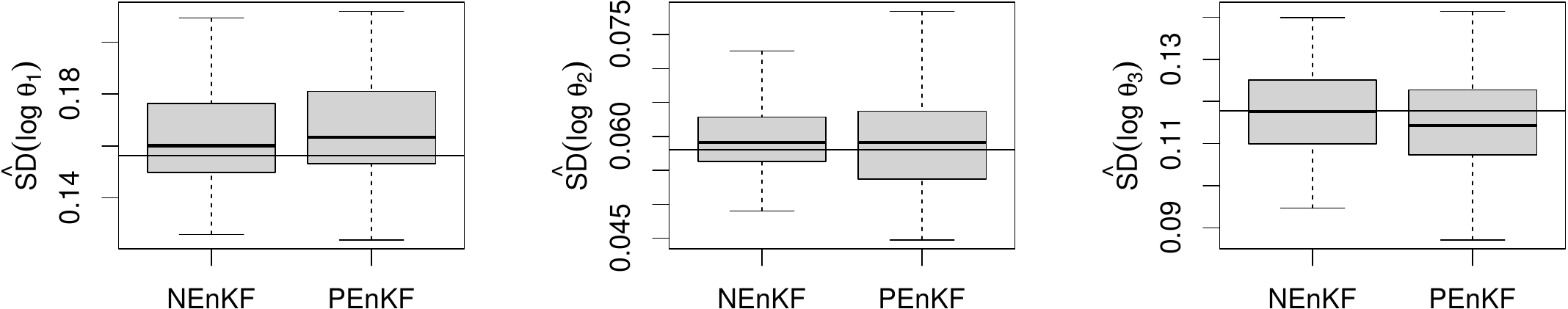}
\caption{OU example. Distributions of the estimators of posterior expectations $\text{E}(\log\theta_i)$ and standard deviations $\text{SD}(\log\theta_i)$, $i=1,2,3$, under NEnKF with $M=1000$ and PEnKF with $M=4000$. The ground truth is indicated by a horizontal line.}
\label{fig:OUest}
\end{figure}

\begin{figure}[ht]
\centering
\includegraphics[width=16.5cm,height=5.3cm]{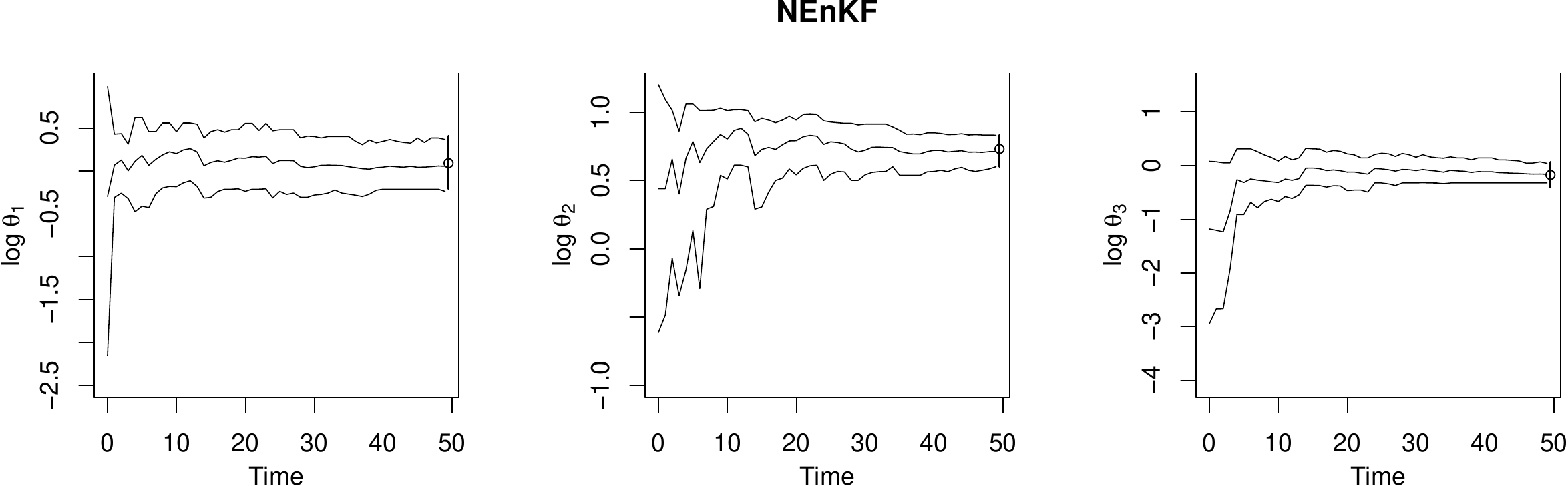}

\

\includegraphics[width=16.5cm,height=5.3cm]{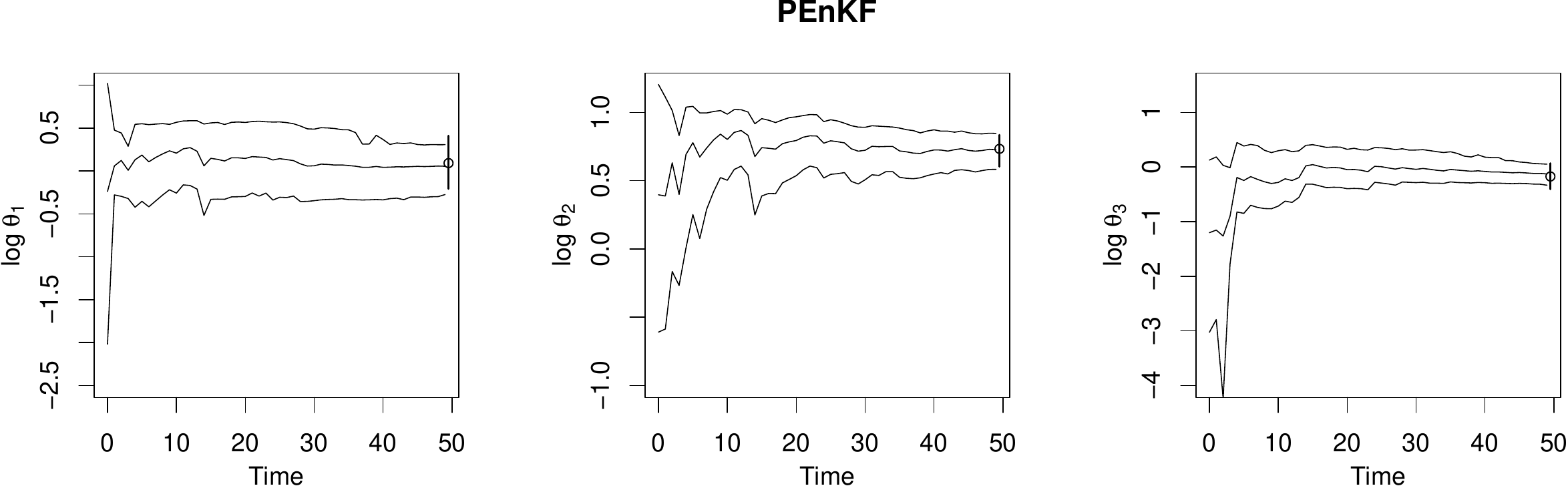}

\

\includegraphics[width=16.5cm,height=5.3cm]{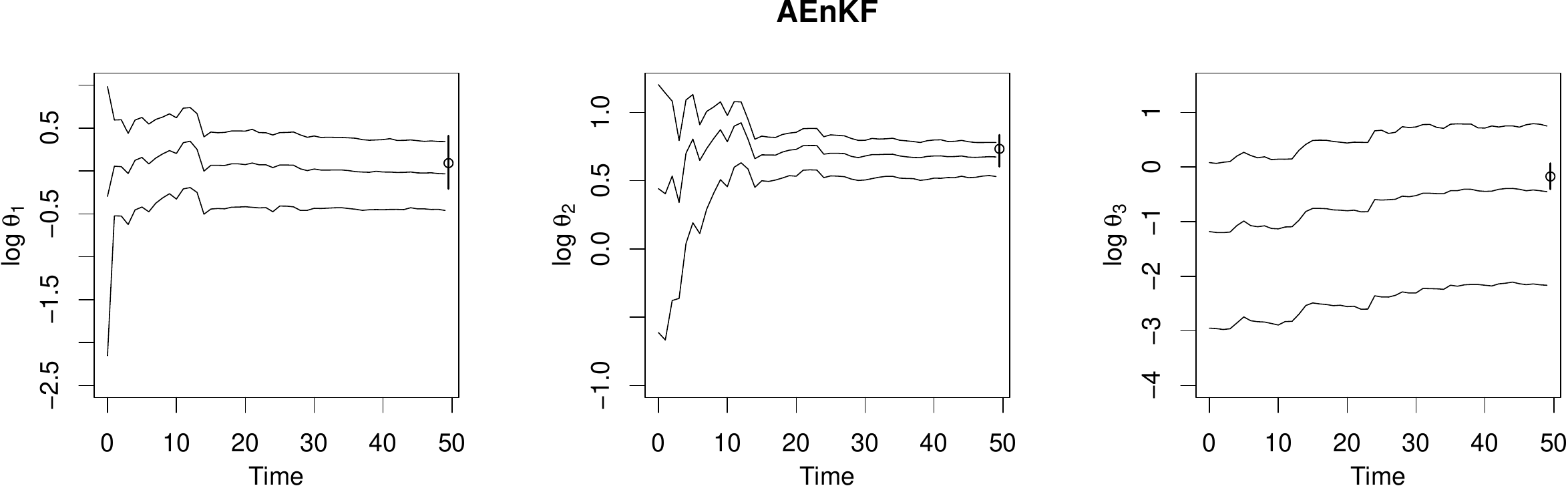}
\caption{OU example. Marginal parameter filtering means and 95\% credible intervals under the NEnKF, PEnKF and AEnKF schemes. For reference, ground truth summaries obtained from the output of MCMC are displayed at time 49.}
\label{fig:OUseq}
\end{figure}

Using the settings described above, 100 replicate runs of each scheme were performed, and the results summarised in Table~\ref{tab:tabOU} and Figures~\ref{fig:OUest}-\ref{fig:OUseq}. We compare the accuracy of each scheme by reporting bias and root mean square error (RMSE) of the estimators of
the marginal posterior means and standard deviations of each $\log\theta_i$. These summaries, denoted $\text{E}(\log\theta_i)$ and $\text{SD}(\log\theta_i)$ respectively, are reported in Table~\ref{tab:tabOU} and were obtained by comparing estimates from replicate runs to `gold standard' reference values, obtained from a long run ($10^6$ iterations) of an MCMC scheme targeting the marginal parameter posterior $\pi_T(\theta|y_{0:T})$, as given by (\ref{ppost}); for an OU process, the observed-data likelihood $p_T^{\theta}(y_{0:T})$ is tractable.

There are considerable differences in accuracy of AEnKF compared to NEnKF and PEnKF, with the former consistently over-estimating all posterior quantities of interest. This is particularly noticeable for the marginal posterior mean and standard deviation of $\log\theta_3$, which controls the intrinsic stochasticity of the latent process, and for which an assumed linear Gaussian relationship between this parameter and the observation process is likely to be unsatisfactory. There is comparatively little difference between NEnKF and PEnKF, although Figure~\ref{fig:OUest} suggests that lower bias and RMSE is achieved for NEnKF. In the remaining applications, therefore, we focus on comparisons between NEnKF and SMC$^2$ (and where appropriate, RB-SMC$^2$).         

\subsection{Lotka-Volterra model}\label{sec:app2}
The Lotka-Volterra system admits a bivariate SDE giving a simple description of the time-course behaviour of prey ($X_{1,t}$) and predators ($X_{2,t}$). It has been ubiquitously used to benchmark competing inference algorithms \citep[see e.g.][]{Fuchs_2013,ryder21,graham22}. The SDE takes the form
\begin{equation}\label{ItoSDE}
dX_t = a(X_t,\theta)dt + b(X_t,\theta)^{1/2} dW_t
\end{equation}
where
\[
a(X_t,\theta)=\begin{pmatrix}\theta_1 X_{1,t}-\theta_2 X_{1,t}X_{2,t} \\\theta_2 X_{1,t}X_{2,t}-\theta_3 X_{2,t} \end{pmatrix},\qquad b(X_t,\theta)=\begin{pmatrix}\theta_1 X_{1,t}+\theta_2 X_{1,t}X_{2,t}   & -\theta_2 X_{1,t}X_{2,t} \\ -\theta_2 X_{1,t}X_{2,t} &\theta_2 X_{1,t}X_{2,t}+\theta_3 X_{2,t} 
\end{pmatrix}.
\]
Although this SDE cannot be solved analytically, we adopt a time discretisation approach by 
partitioning $[t-1,t]$ into $m+1$ equally spaced times $t=t,0<t,1<\ldots <t,m=t$ with time-step $\Delta t$. We then set $p_t^{(\theta)}(x_t|x_{t-1})=\int \prod_{i=1}^m p_{t,i}^{(\theta)}(x_{t,i}|x_{t,i-1})dx_{t,1:m-1}$ where $p_{t,i}^{(\theta)}(x_{t,i}|x_{t,i-1})$ is the Euler-Maruyama approximation:
\begin{equation}\label{em}
p_{t,i}^{(\theta)}(x_{t,i}|x_{t,i-1})=\mathcal{N}\left(x_{t,i}; x_{t,i-1}+\alpha(x_{t,i-1},\theta)\Delta t, \beta(x_{t,i-1},\theta)\Delta t\right).    
\end{equation}
To create a challenging data-poor scenario, we set $X_0=(50,50)'$, $\theta=(0.5,0.0025,0.3)'$ and recursively sampled (\ref{em}) over $[0,40]$ with $\Delta t=10^{-3}$. The resulting trace was thinned by a factor of 2000, to give 20 observations with equal time spacing of 2 time units. We then discarded predator values and corrupted the remaining prey values according to an observation model of the form  
\begin{equation}\label{eqn:lvObs}
{f}_t^{(\theta)}(y_t|x_t)= \mathcal{N}\left(y_t;H x_t,H x_t \right)
\end{equation}
where $H = (1, 0)$. This (nonlinear) observation model effectively represents a Gaussian approximation to a Poisson model, and has been used in application in population ecology by \cite{lowe23} among others. Hence, in addition to vanilla SMC$^2$ (see Section~\ref{subsec:smc2}), we consider use of the EnKF as descsribed in Section~\ref{NGobs}, as appropriate for nonlinear observation models. In more detail, we implemented three schemes as follows:
\begin{enumerate}
\item \emph{Nested EnKF} (NEnKF). The target at time $t$ is $\widehat{\pi}_{\text{enkf},t}(\theta|y_{0:t})$. Observed-data likelihood contributions $\widehat{p}_{\text{enkf},t}^{(\theta)}(y_t|y_{0:t-1},\tilde{x}_t^{1:N})$ are calculated using Algorithm~\ref{alg:EnKF} with $R$ replaced by $\tilde{R}=H\hat{\mu}_{t|t-1}$, permitting a linear Gaussian approximation of (\ref{eqn:lvObs}).  
\item \emph{Rao-Blackwellised SMC$^2$} (RB-SMC$^2$). The target at time $t$ is $\widetilde{\pi}_{\text{enkf},t}(\theta|y_{0:t})$. Estimates of observed-data likelihood contributions $\widetilde{p}_{\text{enkf},t}^{(\theta)}(y_t|y_{0:t-1},\tilde{x}_t^{1:N})$ are calculated using Algorithm~\ref{alg:PartRB} 
with $g_t^{(\theta)}(y_t|x_t)=\mathcal{N}(y_t,H x_t,\tilde{R})$.
To mitigate against a proposal distribution (for generating $x_t^{j}$ in Algorithm~\ref{alg:PartRB}) that is potentially light-tailed compared to the target (\ref{eq:filtNL}), we further scaled $\tilde{R}=H\hat{\mu}_{t|t-1}$ by a factor of 2. 
\item \emph{Vanilla SMC$^2$} (SMC$^2$). The target at time $t$ is $\pi_{t}(\theta|y_{0:t})$. Estimates of observed-data likelihood contributions $p_{t}^{(\theta)}(y_t|y_{0:t-1})$ are calculated using Algorithm~\ref{alg:PartF}, with step 1. replaced by sampling $x_t^{j}\sim p_t^{(\theta)}(\cdot|x_{t-1}^{a_{t-1}^j})$ via recursive application of the Euler-Maruyama discretisation. The weight function then becomes $\tilde{w}_{t}(x_{t-1}^{a_{t-1}^j},x_t^{j})={f}_t^{(\theta)}(y_t|x_t^{j})$. 
\end{enumerate}
An independent prior specification with $\theta_1\sim \text{Gamma}(2,4)$, $\theta_2\sim \text{Gamma}(20,10^4)$ and $\theta_3\sim \text{Gamma}(2,4)$ was adopted. We then ran the above schemes with $M=1000$ particles and initial ensemble members $N=20$. Sampling from $p_t^{(\theta)}(x_t|x_{t-1})$ used (\ref{em}) with $\Delta t =0.2$. The conditions for triggering a resample-move step and increasing $N$ were as outlined in Section~\ref{sec:app1}. All schemes used a delayed acceptance kernel in the move step, based on the $k$-nearest neighbours surrogate of Section~\ref{da}, with $k=3$. This gave a decrease in CPU time of around 30\% over the corresponding schemes which used a standard (pseudo-marginal) MH kernel.    

\begin{table}[t]
\centering
\small
	\begin{tabular}{@{}l lll lllll@{}}
         \toprule
Scheme  & $N$ &  CPU & \multicolumn{6}{l}{Bias (RMSE)}  \\
\cmidrule(l){4-9}
	&         &            & $\widehat{\text{E}}(\log\theta_1)$ & $\widehat{\text{E}}(\log\theta_2)$ & $\widehat{\text{E}}(\log\theta_3)$ &$\widehat{\text{SD}}(\log\theta_1)$ & $\widehat{\text{SD}}(\log\theta_2)$ & $\widehat{\text{SD}}(\log\theta_3)$\\   
\midrule
NEnKF     &20             & 129           &0.0033 &-0.0128 &-0.0201 &-0.0021 &-0.0020 &-0.0009  \\
 & & & (0.019) & \phantom{-}(0.024) & \phantom{-}(0.029) & \phantom{-}(0.009) & \phantom{-}(0.011) & \phantom{-}(0.011) \\
RB-SMC$^2$&           40  &            287&0.0040 &-0.0089 &-0.0097 &-0.0054 &-0.0058 &-0.0042  \\
 & & & (0.021) & \phantom{-}(0.020) & \phantom{-}(0.021) & \phantom{-}(0.012) & \phantom{-}(0.013) & \phantom{-}(0.014) \\
SMC$^2$   &            88 &            561&\phantom{-}0.0047 &-0.0015 &\phantom{-}0.0024 &-0.0073 &-0.0066 &-0.0033 \\
 & & & (0.016) & \phantom{-}(0.020) & \phantom{-}(0.024) & \phantom{-}(0.009) & \phantom{-}(0.011) & \phantom{-}(0.012) \\
\bottomrule
\end{tabular}
	\caption{LV example. Number of ensemble members $N$ (at termination), CPU time (in seconds), bias (and RMSE in parentheses) of estimators 
of the posterior expectations $\text{E}(\log\theta_i)$ and standard deviations $\text{SD}(\log\theta_i)$, $i=1,2,3$. All results are obtained by averaging over 50 runs of each inference scheme with $M=1000$ parameter particles.}\label{tab:tabLV}
\end{table}

\begin{figure}[ht]
\centering
\includegraphics[width=15.5cm,height=3.9cm]{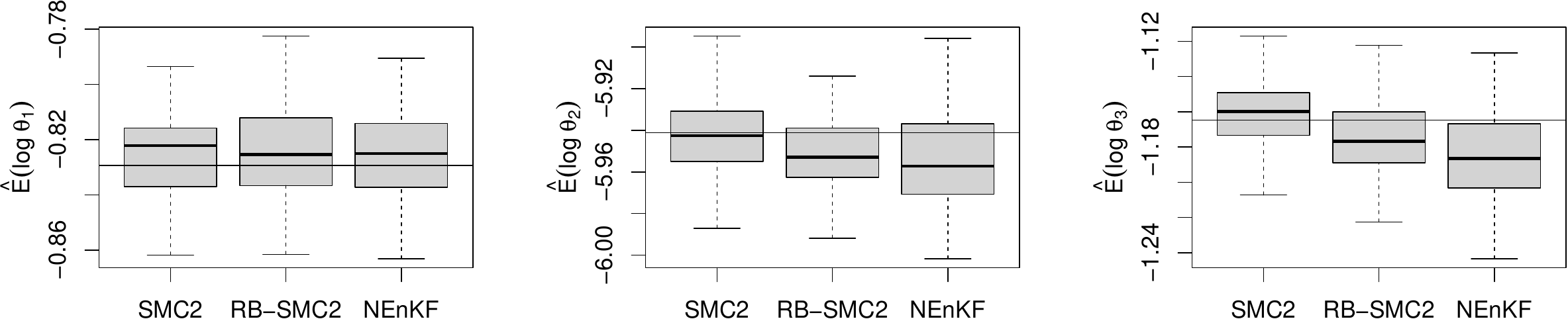}

\

\includegraphics[width=15.5cm,height=3.9cm]{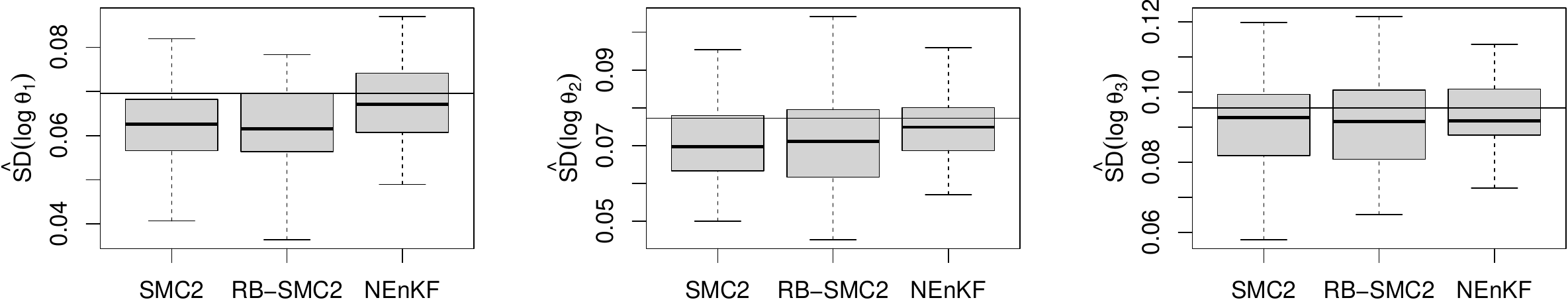}
\caption{LV example. Distributions of the estimators of posterior expectations $\text{E}(\log\theta_i)$ and standard deviations $\text{SD}(\log\theta_i)$, $i=1,2,3$, SMC$^2$, RB-SMC$^2$ and NEnKF, each with $M=1000$. The ground truth is indicated by a horizontal line.}
\label{fig:LVest}
\end{figure}


Using the settings described above, 50 replicate runs of each scheme were performed, and the
results summarised in Table~\ref{tab:tabLV} and Figure~\ref{fig:LVest}. 
We again compare the accuracy of each scheme by reporting bias and root mean square error (RMSE) of the estimators of
the marginal posterior means and standard deviations of each $\log\theta_i$. For this application, we computed reference values of these quantities from a long run ($10^6$ iterations with $100$ state particles) of particle MCMC \citep[see \emph{e.g.}][for an application of this scheme to the Lotka-Volterra SDE]{GoliWilk11}. Hence, for this application, we report accuracy with reference to the `ground truth' marginal parameter posterior $\pi_{T}(\theta|y_{0:T})$, which is also the target of vanilla SMC$^2$, albeit using a particle filter. 

Table~\ref{tab:tabLV} reports, \emph{inter alia}, average run times and the average number of ensemble members at termination of each scheme. NEnKF is around a factor of 4 quicker than vanilla SMC$^2$ and around a factor of 2 quicker than RB-SMC$^2$, with the latter twice as fast as vanilla SMC$^2$. However, NEnKF uses a misspecified variance: recall that NEnKF uses the forecast mean to approximate the observation variance $Hx_t$ at time $t$. Consequently, there are some substantive differences in accuracy between the NEnKF output and gold standard posterior reference values. From Figures~\ref{fig:LVest}, it is clear that for NEnKF, estimates of the parameter posterior means are shifted downward for $\theta_2$ and $\theta_3$; there is relatively little difference in estimated posterior standard deviations for each scheme. RB-SMC$^2$ is able to alleviate these differences in estimated posterior means, giving estimated posterior quantities that are broadly consistent with reference values, albeit with greater computational effort than NEnKF. 

\subsection{Two-node epidemic model}\label{sec:app2b}
We consider here a two-node compartmental epidemic model to characterise the spread of oak processionary moth (OPM) in two London parks (Bushy Park and Richmond Park). Similar to  \cite{wadkin23}, we specify a stochastic susceptible-infected-removed (SIR) model \citep[\emph{e.g.}][]{AnBr00} for each node (park), with common removal rate $\gamma$ and infestation `pressure' from the neighbouring node, described by the parameters $\alpha_{ij}$, where $\alpha_{12}$ is the pressure applied by node 1 (Bushy) on node 2 (Richmond), and $\alpha_{21}$ is the pressure from node 2 on node 1. Unlike \cite{wadkin23}, we allow time-varying infestation rates $\beta_{1,t}$ and $\beta_{2,t}$ at each node, described by two independent Brownian motion processes, scaled by a common parameter $\sigma_{\beta}$. The model takes the form of (\ref{ItoSDE}), with $X_t=(S_{1,t},I_{1,t},S_{2,t},I_{2,t},\log\beta_{1,t},\log\beta_{2,t})'$ and $\theta=(\gamma,\alpha_{1,2},\alpha_{2,1},\sigma_{\beta})'$. The drift and diffusion functions are given by
\[
a(X_t,\theta)=\begin{pmatrix} -\beta_{1,t}(I_{1,t}+\alpha_{21}I_{2,t})S_{1,t}\\
\beta_{1,t}(I_{1,t}+\alpha_{21}I_{2,t})S_{1,t}-\gamma I_{1,t}\\
-\beta_{2,t}(I_{2,t}+\alpha_{12}I_{1,t})S_{2,t}\\
\beta_{2,t}(I_{2,t}+\alpha_{12}I_{1,t})S_{2,t}-\gamma I_{2,t}\\
0\\0\end{pmatrix}, \qquad b(X_t,\theta)= \text{diag}\{b_{11},b_{22},b_{33}\}
\]
where
\[
b_{11}=\begin{pmatrix}\beta_{1,t}(I_{1,t}+\alpha_{21}I_{2,t})S_{1,t}& -\beta_{1,t}(I_{1,t}+\alpha_{21}I_{2,t})S_{1,t}\\
-\beta_{1,t}(I_{1,t}+\alpha_{21}I_{2,t})S_{1,t} & \beta_{1,t}(I_{1,t}+\alpha_{21}I_{2,t})S_{1,t}+\gamma I_{1,t}\end{pmatrix},
\]
\[
b_{22}= \begin{pmatrix}\beta_{2,t}(I_{2,t}+\alpha_{12}I_{1,t})S_{2,t} & -\beta_{2,t}(I_{2,t}+\alpha_{12}I_{1,t})S_{2,t}\\ 
-\beta_{2,t}(I_{2,t}+\alpha_{12}I_{1,t})S_{2,t} & \beta_{2,t}(I_{2,t}+\alpha_{12}I_{1,t})S_{2,t}+\gamma I_{2,t}\end{pmatrix}
\]
and $b_{33}=\sigma^2_{\beta}$. 

\begin{figure}[ht]
\centering
\includegraphics[width=14.0cm,height=4.8cm]{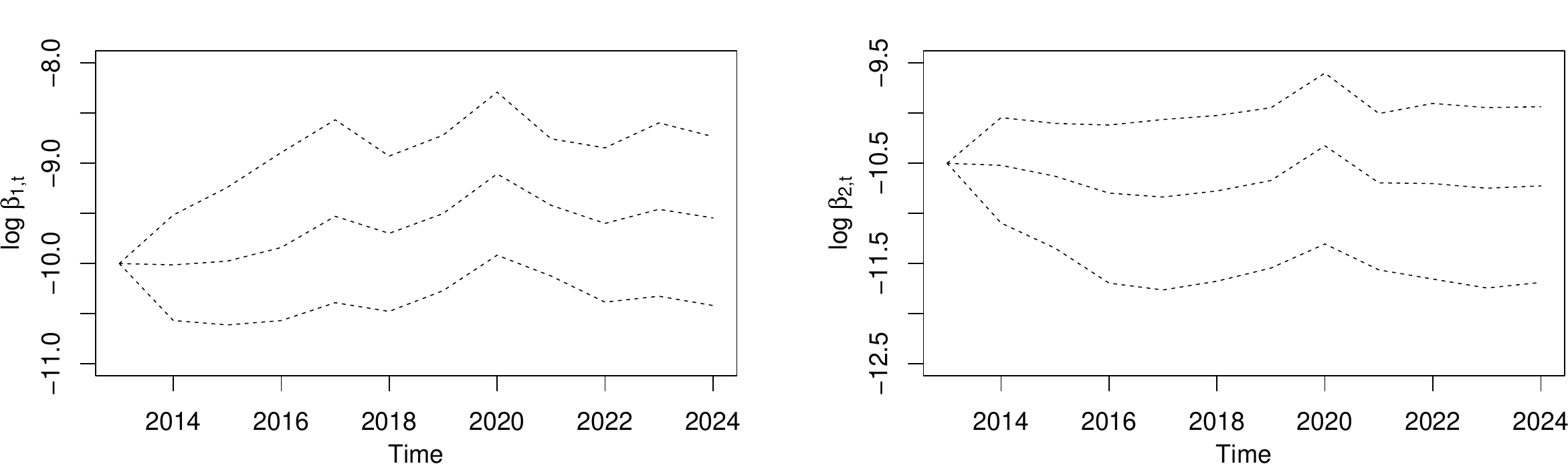}
\includegraphics[width=14.0cm,height=4.8cm]{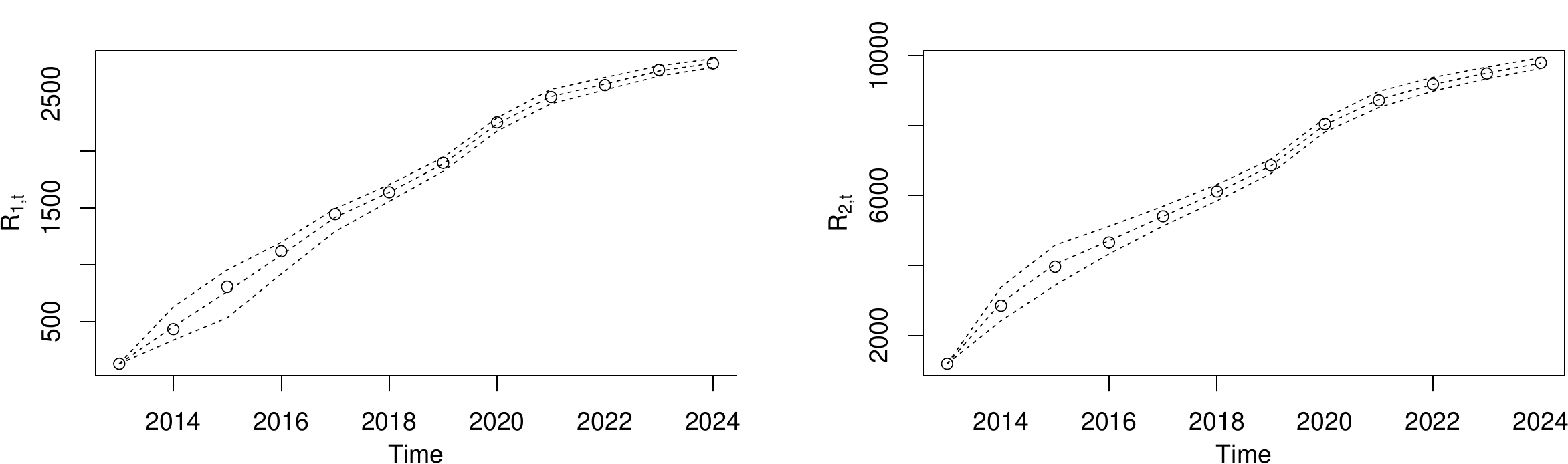}
\caption{Two-node epidemic example. Marginal filtering means and 95\% credible intervals for $\log\beta_{1,t}$ and $\log\beta_{2,t}$ (top panel), and removal prevalences $R_{1,t}$ and $R_{2,t}$ (bottom panel) using the output of NEnKF. Observations are shown as circles.}
\label{fig:SIR}
\end{figure}

The data (see Figure~\ref{fig:SIR}) consist of (noisy) observations of removal prevalence $(R_{1,t},R_{2,t})'$, that is, the cumulative numbers of tree locations where nest removal has taken place, observed each year over the period 2013-2024. We assume a fixed number of tree locations ($\text{N}_j$) at each node $j$ so that $S_{j,t}+I_{j,t}+R_{j,t}=\text{N}_j$, $j=1,2$, and observation of removals is equivalent to observation of the sums $S_{1,t}+I_{1,t}$ and $S_{2,t}+I_{2,t}$. Following \cite{wadkin23}, we specify an observation model with density
\begin{equation}\label{obsSIR}
{f}_t^{(\theta)}(y_t|x_t)= \mathcal{N}\left(y_t;H x_t,\sigma^2 \text{diag}\{H x_t\} \right)
\end{equation}
where
\[
H=\begin{pmatrix}1 & 1 & 0 & 0 & 0 & 0\\
0 & 0 & 1 & 1 & 0 & 0\end{pmatrix}.
\]
We assume that $\sigma$ is unknown and append it to the parameter vector $\theta$. 

We adopted an independent prior specification with $\text{Gamma}(2,2)$, for all parameter components except $\sigma_{\beta}$, for which we ascribed a $\text{Gamma}(2,10)$ distribution, reflecting our prior belief that the intrinsic stochasticity in the infestation process is small, relative to the observation noise. We ran NEnKF and RB-SMC$^2$, which use Algorithm~\ref{alg:EnKF} and \ref{alg:PartRB} respectively, and require a linear Gaussian approximation of (\ref{obsSIR}). For NEnKF, we used $\tilde{R}=\sigma^2 \text{diag}\{H \hat{\mu}_{t|t-1}\}$; for RB-SMC$^2$, this quantity was further scaled by a factor of 2. We performed a single run of these schemes with $M=20,000$ particles and $N=20$ initial ensemble members. An initial condition of $x_0=(4631, 240, 37413, 1400, -10, -10.5)'$ was assumed to be fixed and known for each run. We followed Section~\ref{sec:app2} by sampling $p_t^{(\theta)}(x_t|x_{t-1})$ through recursive draws of the Euler-Maruyama approximation  (\ref{em}) with $\Delta t =0.1$. The conditions for triggering a resample-move step and increasing $N$ were as Section~\ref{sec:app1}. Both schemes used a delayed acceptance kernel in the move step (with $k=3$), which gave a decrease in CPU time of approximately 15\% for NEnKF and 25\% for RB-SMC$^2$, over use of a standard (pseudo-marginal) MH kernel.

\begin{figure}[ht]
\centering
\includegraphics[width=16.0cm,height=4.0cm]{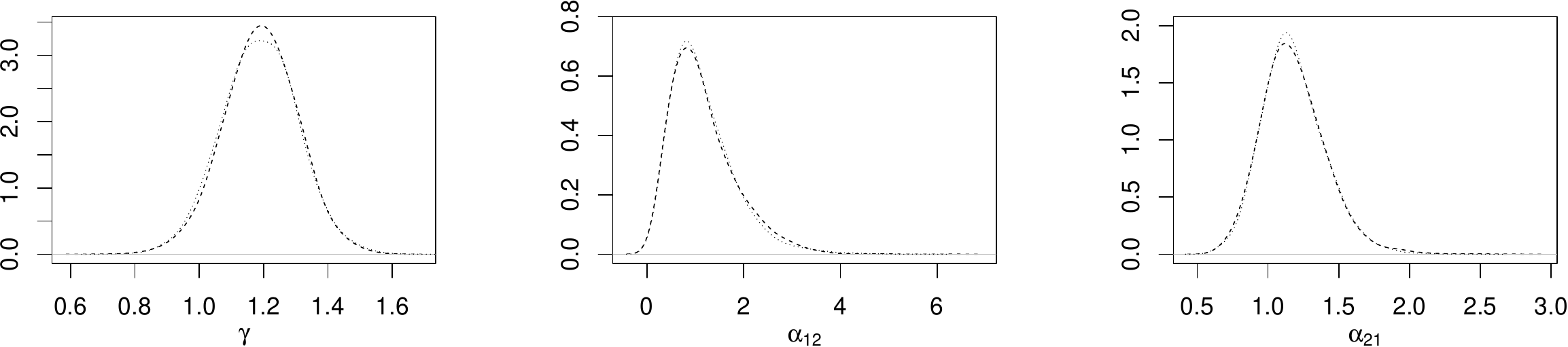}
\includegraphics[width=11.0cm,height=4.0cm]{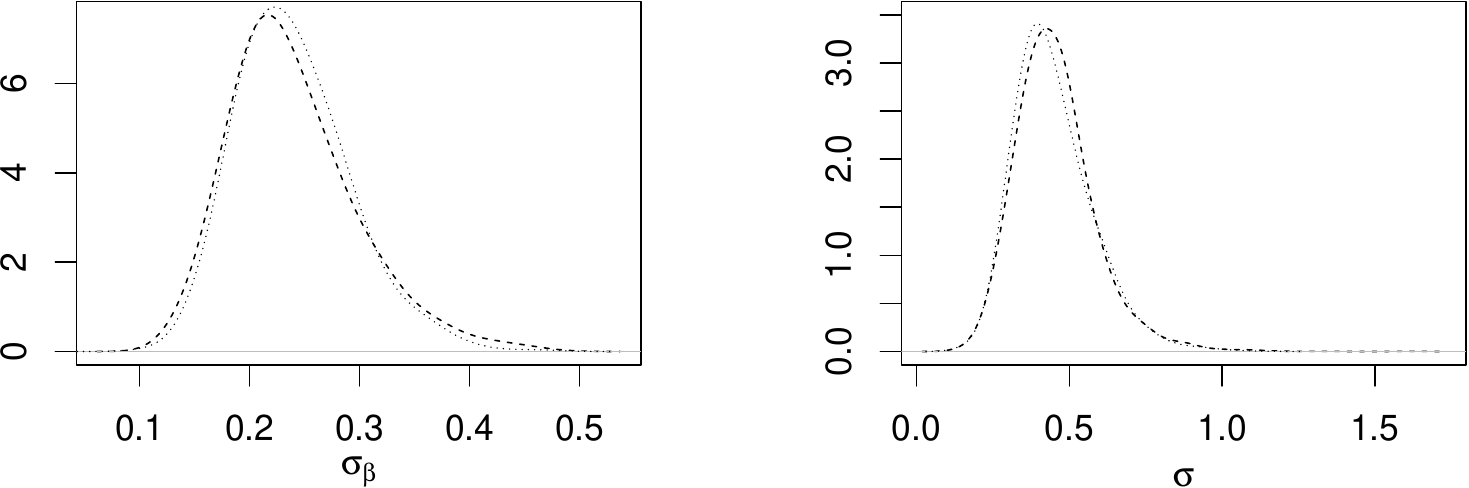}
\caption{Two-node epidemic example. Marginal parameter posterior distributions at time $T$ from the output of RB-SMC$^2$ (dotted line) and NEnKF (dashed line) using $M=5\times 10^4$.}
\label{fig:SIRdens}
\end{figure}

The output of NEnKF and RB-SMC$^2$ is summarised in Figures~\ref{fig:SIRdens} and \ref{fig:SIR}. The former shows good agreement in posterior output from both schemes while the latter shows that the model can generate predicted removal prevalences that are consistent with the data, and that the infestation rates are plausibly constant. NEnKF terminated with $N=85$ ensemble members, whereas RB-SMC$^2$ terminated with $N=300$, giving a relative CPU cost of roughly 1:2.7. It is worth noting here that computational resources precluded running vanilla SMC$^2$. A typical termination value of $N$ for SMC$^2$ can be determined by requiring $\text{Var}[\widehat{p}_{T}^{(\bar{\theta})}(y_{0:T}|x_{0:T}^{1:N},a_{0:T-1}^{1:N})]\approx 1.5$, where $\bar{\theta}$ is the parameter posterior mean estimated from RB-SMC$^2$. This gave $N\approx 5,000$, underlining the benefit of propagating the latent state process conditional on the next observation (as in NEnKF and RB-SMC$^2$) as opposed to myopically (as in vanilla SMC$^2$). 

\subsection{Lorenz 96}\label{sec:app3}
Consider a process $\{X_t,t\geq 0\}$ satisfying the It\^o SDE
\begin{equation}\label{LorSDE}
dX_t = a(X_t,\theta)dt + \Sigma^{1/2} dW_t.
\end{equation}
Here, the drift $a(X_t,\theta)$ is a length-$d$ vector with $i$th component
\[
a_i(X_t,\theta)=\theta_1(X_{i+1,t}-X_{i-2,t})X_{i-1,t}-\theta_2 X_{i,t}+\theta_3
\]
with addition and subtraction of component indices interpreted modulo $d$. The diffusion matrix is a diagonal matrix such that $\Sigma=\theta_4^2 I_{d\times d}$. This SDE can be seen as a stochastic representation of the Lorenz 96 dynamical system \citep{Lorenz1996}, and is often used for benchmarking inference schemes in high-dimensional settings. Following \cite{drovandi22}, we set $x_0=(0,\ldots,0)'$ and $\theta=(1,1,8,\sqrt{10})'$. Two synthetic data sets, $\mathcal{D}_1$ (with $d=5$) and $\mathcal{D}_2$ (with $d=10$), were generated via recursive sampling of the Euler-Maruyama approximation of (\ref{LorSDE}) over $[0,30]$ with $\Delta t=5\times 10^{-3}$. The resulting paths were thinned by a factor of 40 and corrupted with additive Gaussian noise, with zero mean and variance $R=5^2 I_{d\times d}$, giving, for each of $\mathcal{D}_1$ and $\mathcal{D}_2$, 30 observations with an inter-observation time of $0.2$ time units.  

We adopted an independent prior specification with $\theta_1\sim \text{Gamma}(4,4)$, $\theta_2\sim \text{Gamma}(4,4)$, $\theta_3\sim \text{Gamma}(6,2)$ and $\theta_4\sim \text{Gamma}(16,2)$. We implemented NEnKF and SMC$^2$ with $M=1000$ particles and initial ensemble members $N=20$ for data set $\mathcal{D}_1$ and $N=50$ for data set $\mathcal{D}_2$. Sampling from $p_t^{(\theta)}(x_t|x_{t-1})$ used the Euler-Maruyama approximation of (\ref{LorSDE}) with $\Delta t =5\times 10^{-3}$. The conditions for triggering a resample-move step and increasing $N$ were as outlined in Section~\ref{sec:app1}. Both schemes used a delayed acceptance kernel in the move step, based on the $k$-nearest neighbours surrogate of Section~\ref{da}, with $k=3$. We found that performing 5 iterations of the move step (per parameter particle) was necessary to give adequate parameter rejuvenation for this particular application, likely due to the calculation of particularly small ESS values at some time points, in turn resulting in a surrogate with relatively poor accuracy for some proposed parameter values.    

\begin{table}[t]
\centering
\small
	\begin{tabular}{@{}l lll lllll@{}}
         \toprule
Scheme  & $N$ &  CPU & \multicolumn{6}{l}{Bias (RMSE)}  \\
\cmidrule(l){4-9}
	&         &            & $\widehat{\text{E}}(\log\theta_1)$ & $\widehat{\text{E}}(\log\theta_2)$ & $\widehat{\text{E}}(\log\theta_3)$ &$\widehat{\text{SD}}(\log\theta_1)$ & $\widehat{\text{SD}}(\log\theta_2)$ & $\widehat{\text{SD}}(\log\theta_3)$\\   
\midrule
NEnKF     &128            & 14     &0.2082 &-0.1804 &0.1494 &0.0759 &0.1465 &-0.0064  \\
          &               &          &(0.263) &\phantom{-}(0.277) &(0.120) &(0.105) &(0.149) & \phantom{-}(0.043) \\
SMC$^2$   &450            & 56     &0.2109 &-0.1709 &0.1158 &0.0500 &0.0816 &\phantom{-}0.0084  \\
          &               &          &(0.170) & \phantom{-}(0.205) &(0.098) &(0.044) &(0.109) & \phantom{-}(0.043) \\
\bottomrule
\end{tabular}
	\caption{Lorenz 96 example. Number of ensemble members $N$ (at termination), CPU time (in minutes), bias (and RMSE in parentheses) of estimators 
of the posterior expectations $\text{E}(\log\theta_i)$ and standard deviations $\text{SD}(\log\theta_i)$, $i=1,2,3$. All results are obtained by averaging over 50 runs of each inference scheme with $M=1000$ parameter particles using data set $\mathcal{D}_1$ ($d=5$).}\label{tab:tabLorenz}
\end{table}

\begin{figure}[ht]
\centering
\includegraphics[width=15.5cm,height=4.0cm]{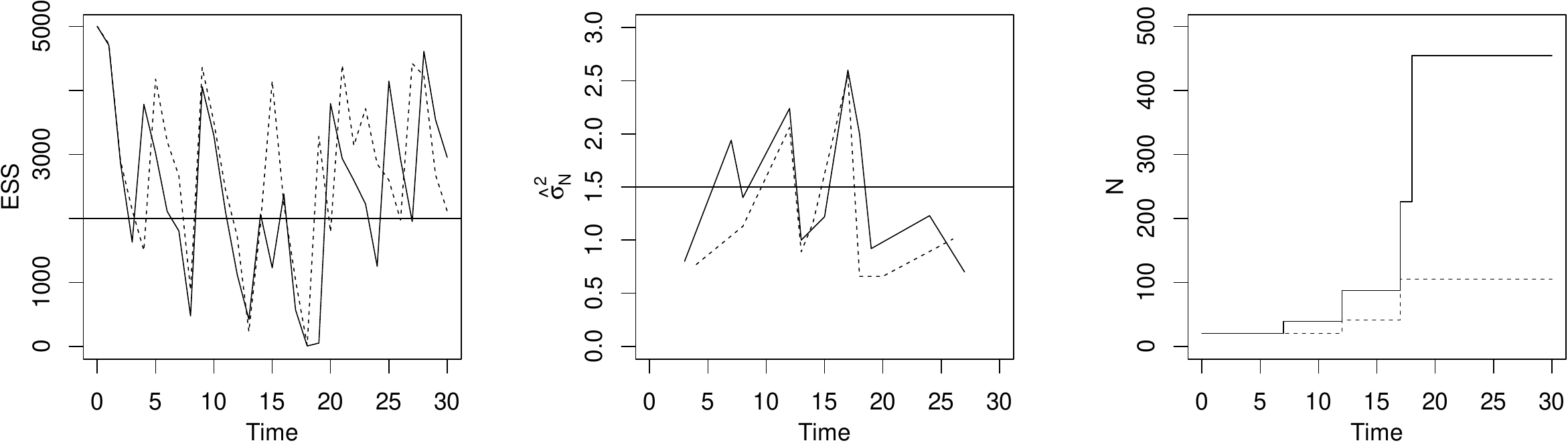}

\

\includegraphics[width=15.5cm,height=4.0cm]{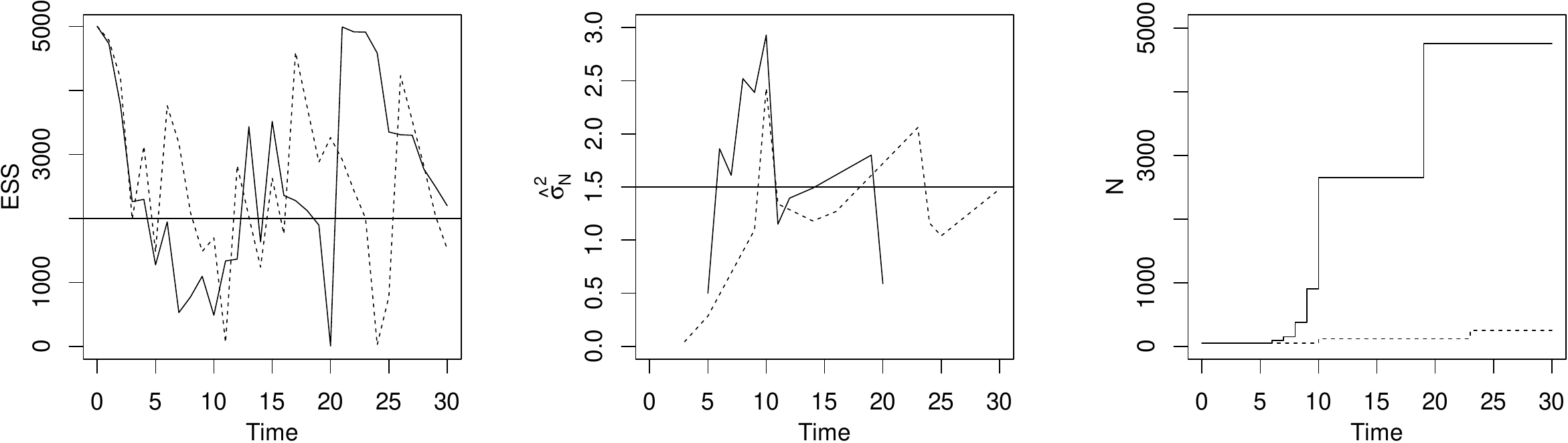}
\caption{Lorenz 96 example. Effective sample size (ESS) against time (left), $\hat{\sigma}^2_N$ 
against time and number of state particles / ensemble members $N$ against time. Horizontal lines indicate
the thresholds at which resampling and increasing $N$ take place. All results are based on a single
typical run of SMC$^2$ (solid lines) and NEnKF (dashed lines) using data set $\mathcal{D}_1$ (top panel) and data set $\mathcal{D}_2$ (bottom panel).}
\label{fig:LOR1}
\end{figure}

\begin{figure}[ht]
\centering
\includegraphics[width=14.0cm,height=4.8cm]{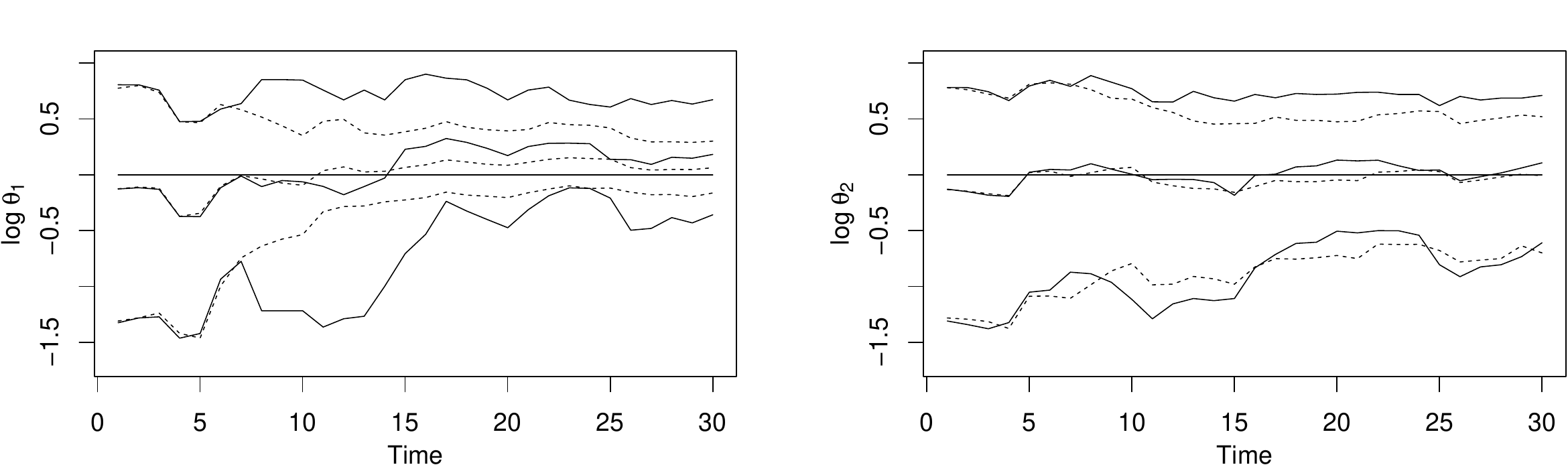}
\includegraphics[width=14.0cm,height=4.8cm]{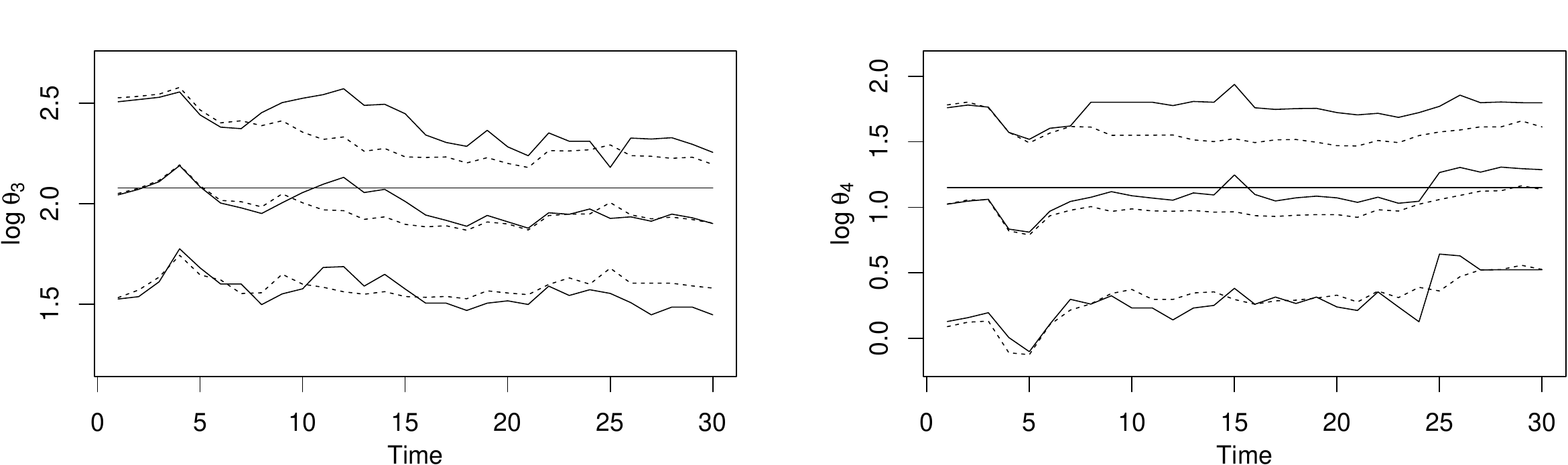}
\caption{Lorenz 96 example. Marginal parameter filtering means and 95\% credible intervals under SMC$^2$ (solid lines) and NEnKF (dashed lines). For reference, ground truth parameter values are displayed as horizontal lines.}
\label{fig:LOR2}
\end{figure}

Using the settings described above, 50 replicate runs of both schemes were performed for data set $\mathcal{D}_1$, and the
results summarised in Table~\ref{tab:tabLorenz}, with accuracy of various posterior estimators compared to reference values computed from particle MCMC ($10^6$ iterations with $300$ state particles). We were unable to repeat this process for data set $\mathcal{D}_2$; in particular running SMC$^2$ and particle MCMC require around $4,000-5,000$ state particles, precluding replicate runs of SMC$^2$ and particle MCMC. For this data set, we compare efficiency of SMC$^2$ and NEnKF via a single run with $M=5,000$ parameter particles; see Figures~\ref{fig:LOR1}--\ref{fig:LOR2}. 

Inspection of Table~\ref{tab:tabLorenz} and Figure~\ref{fig:LOR2} suggests that the NEnKF is able to give inferences consistent with SMC$^2$ and the ground truth parameter values that generated the data. For data set $\mathcal{D}_1$ ($d=5$), fewer ensemble members are required for NEnKF compared to state particles for SMC$^2$, leading to an increase in computational efficiency of around a factor of 3.5. For data set $\mathcal{D}_2$ ($d=10$) and a single run of each scheme, SMC$^2$ required 4762 state particles whereas NEnKF required just 249 ensemble members, leading to a difference in computational efficiency of approximately a factor of 19 (2737 minutes versus 147 minutes, for SMC$^2$ versus NEnKF). Not surprisingly, as the observation dimension increases, both schemes require increased numbers of state particles / ensemble members, $N$, although the relative increase is much smaller when using NEnKF. It is evident from Figure~\ref{fig:LOR2}, that NEnKF triggers the resample-move and dynamic increasing of $N$ steps less often than SMC$^2$. When the latter is triggered, realisations of $\hat{\sigma}_N^2$ are typically smaller for NEnKF than SMC$^2$.

\section{Discussion}
\label{sec:disc}
Bayesian filtering for state-space models in the presence of unknown static parameters is a challenging problem. Sequential Monte Carlo algorithms must overcome particle degeneracy; one such scheme that deals with static parameters in a principled way is the SMC$^2$ scheme of \cite{Chopin2013}. Here, parameter particles are weighted by an estimate of the current observed-data likelihood and, subject to some degeneracy criterion, mutated via a (pseudo-marginal) Metropolis-Hastings kernel which has the marginal parameter posterior as its invariant distrubution. A particle filter over the latent state process is used to estimate the required observed-data likelihood contributions. However, use of the particle filter in this way is likely computationally prohibitive in high-dimensional observation settings, due to requiring that the number of state particles be scaled exponentially with dimension. On the other hand, the ensemble Kalman filter (EnKF, \citealp{evensen1994}) shifts particles (known as ensemble members in the EnKF context) via suitable Gaussian approximations, thus avoiding resampling altogether. Consequently, there is now a vast literature demonstrating the empirical performance of the EnKF for high-dimensional state-space models \citep[see \emph{e.g.}][and the references therein]{Katzfuss2019}.

We have proposed to use the EnKF within the principled framework of SMC$^2$ by weighting parameter particles by the EnKF likelihood and, when necessary, mutating these particles by using a Metropolis-Hastings step that again uses the EnKF likelihood. Extensions to the resulting \emph{nested EnKF (NEnKF)} algorithm were considered. In particular, the computational cost of the mutation step can be lessened by screening proposed moves using a surrogate, computed locally using the most recent parameter samples and their likelihood evaluations.  As with the simplest implementation of SMC$^2$, NEnKF is plug-and-play, since only forward simulation of the latent state process and evaluation of the observation density are required. We found, across four applications of increasing observation dimension, that the NEnKF is able to provide accurate and computationally efficient inferences regarding the static parameters. Since the NEnKF combines a particle filter over parameters with an ensemble Kalman filter over states, we anticipate (and as evidenced by our numerical experiments) greatest utility of the proposed algorithm in settings where the number of observed dynamic components is large compared to the number of parameters to be estimated (\emph{e.g.} $d_{\theta}<10$ and $d_y>10$). The simplest implementation of the NEnKF assumes a linear Gaussian observation model. In scenarios where this assumption is unreasonable, we have proposed to replace the EnKF with a particle filter that uses the EnKF to propagate state particles. Unlike related approaches \citep[\emph{e.g.}][]{papadakis2010}, we also use the EnKF to give a linear Gaussian approximation of the transition density $p_t^{(\theta)}(x_t|x_{t-1})$, leading to a simple form of weight function that can be readily applied in settings where states are modelled at a finer frequency than observations (as is the case, for example, when working with discretised SDEs). The resulting Rao-Blackwellised particle filter can be used as the inner filter in SMC$^2$, giving an inference scheme that is typically less accurate but more efficient than vanilla SMC$^2$, and less efficient but more accurate than NEnKF.     

This work can be extended in a number of ways. For example, in our applications of the delayed acceptance kernel, the number of distinct parameter particles $M_r$ used to construct the $k$-nn surrogate was typically small (no more than a few thousand particles), justifying an $\mathcal{O}(M_r)$ list look-up method for finding the $k$-nearest neighbours of each $\theta^*$. Using a variation of the KD-tree \cite[\emph{e.g.}][]{Sherlock2017} requires $\mathcal{O}(d_{\theta}\log M_r)$ operations and provides a cheaper alternative to the list look-up method, provided $d_\theta$ is moderate. It may also be possible to improve computational efficiency of the NEnKF by adaptively choosing between several delayed-acceptance kernels (with different $k$) and a standard M-H kernel, when the resample-move step is triggered. For example, in the early phase of the NEnKF scheme, calculating an observed-data likelihood estimator is likely to be relatively cheap, motivating use of a standard M-H kernel. Similar approaches have been considered in the context of IBIS \citep{taylor13} and SMC$^2$ \citep{botha23adapt}. The efficiency of the resample-move step could be further improved by leveraging gradient-based proposals in the M-H step \cite[e.g.][]{rosato2024}. Finally, using the surrogate in both the weighting and resample-move steps (for a fixed value of $N$) could provide a fixed cost alternative to the NEnKF, to the detriment of inferential accuracy.

\bibliographystyle{apalike}
\bibliography{Refs}



\end{document}